\documentclass[10pt]{article}

\usepackage{amsmath}
\usepackage{amssymb}

\usepackage{graphicx}

\usepackage{cite}

\usepackage{color}

\newcommand{\Prob}{\mathbb{P}}
\newcommand{\R}{\mathbb{R}}
\newcommand{\E}{\mathbb{E}}
\newcommand{\N}{\mathbb{N}}

\newcommand{\T}{\intercal}

\newcommand{\phib}{\mathbf{\emptyset}}
\newcommand{\X}[1]{\mathbf{S_{#1}}}
\newcommand{\Xo}{\mathbf{S}}
\newcommand{\eps}{\varepsilon}

\newcommand{\longrightharpoonup}{\relbar\joinrel\relbar\joinrel\rightharpoonup}
\newcommand{\llongrightharpoonup}{\relbar\joinrel\relbar\joinrel\relbar\joinrel\rightharpoonup}

\def\llongrightarrow{\relbar\joinrel\relbar\joinrel\relbar\joinrel\rightarrow}
\providecommand{\rarrow}[1]{\stackrel{#1}{\llongrightarrow}}

\newtheorem{theorem}{Theorem}[section]
\newtheorem{proposition}[theorem]{Proposition}
\newcommand{\mendex}{\hfill \ensuremath{\diamond}}

\newtheorem{opt}[theorem]{Optimization problem}

\newtheorem{condition}[theorem]{Condition}

\newtheorem{result}[theorem]{Result}

\newcommand{\red}[1]{\textcolor{black}{#1}}

\title{A scalable computational framework for establishing long-term behavior of stochastic reaction networks}

\author{Ankit Gupta, Corentin Briat, Mustafa Khammash$^{*}$\\Department of Biosystems Science and Engineering (D-BSSE),\\Swiss Federal Institute of Technology--Z\"{u}rich (ETH-Z), 4058 Basel, Switzerland\\Corresponding author; e-mail: mustafa.khammash@bsse.ethz.ch}

\begin{document}

\maketitle

\section*{Abstract}

Reaction networks are systems in which the populations of a finite number of species evolve through predefined interactions. Such networks are found as modeling tools in many biological disciplines such as biochemistry, ecology, epidemiology, immunology, systems biology and synthetic biology. It is now well-established that, for small population sizes, stochastic models for biochemical reaction networks are necessary to capture randomness in the interactions. The tools for analyzing such models, however, still lag far behind their deterministic counterparts. In this paper, we bridge this gap by developing a constructive framework for examining the long-term behavior and stability properties of the reaction dynamics in a stochastic setting. In particular, we address the problems of determining ergodicity of the reaction dynamics, which is analogous to having a globally attracting fixed point for deterministic dynamics. We also examine when the statistical moments of the underlying process remain bounded with time and when they converge to their steady state values. The framework we develop relies on a blend of ideas from probability theory, linear algebra and optimization theory. \red{We demonstrate that the stability properties of a wide class of biological networks can be assessed from our sufficient theoretical conditions that can be recast as efficient and scalable  linear programs, well-known for their tractability.} It is notably shown that the computational complexity is often linear in the number of species. We illustrate the validity, the efficiency and the \red{wide applicability} of our results on several reaction networks arising in biochemistry, systems biology, epidemiology and ecology. The biological implications of the results as well as an example of a non-ergodic biological network are also discussed.

\newpage

%
\section*{Author Summary}

In many biological disciplines, computational modeling of interaction networks is the key for understanding biological phenomena. Such networks are traditionally studied using deterministic models. However, it has been recently recognized that when the populations are small in size, the inherent random effects become significant and to incorporate them, a stochastic modeling paradigm is necessary. Hence, stochastic models of reaction networks have been broadly adopted and extensively used. Such models, for instance, form a cornerstone for studying heterogeneity in clonal cell populations.

In biological applications, one is often interested in knowing the long-term behavior and stability properties of reaction networks even with incomplete knowledge of the model parameters. However for stochastic models, no analytical tools are known for this purpose, forcing many researchers to use a simulation-based approach, which is highly unsatisfactory. To address this issue, we develop a theoretical and computational framework for determining the long-term behavior and stability properties for stochastic reaction networks. Our approach is based on a mixture of ideas from probability theory, linear algebra and optimization theory. We illustrate the broad applicability of our results by considering examples from various biological areas. The biological implications of our results are discussed as well.

\section*{Introduction}

Reaction networks are used as modeling tools in many areas of science. Examples include chemical reaction networks \cite{BookCR}, cell signalling networks \cite{CellSignaling}, gene expression networks \cite{MO}, metabolic networks \cite{Schuetz}, pharmacological networks \cite{Berger}, epidemiological networks \cite{Hethcote} and ecological networks \cite{Bascompte}. Traditionally, reaction networks are mathematically analyzed by representing the dynamics as a set of ordinary differential equations. Such a deterministic model is reasonably accurate when the number of network participants is \emph{large}. However, when this is not the case, the discreteness in the interactions becomes important and the dynamics inherently \emph{noisy}. This \emph{random} component of the dynamics cannot be ignored as it can strongly influence the system's  behavior \cite{Goutsias,McAdams,Levin}. To understand the effects of this randomness, stochastic models are needed, and the most common approach is to model the reaction dynamics as a continuous-time Markov process. The most common approach is to model the dynamics as a continuous-time Markov process whose states denote the current population size. Many recent works have employed such stochastic models to study the impact of noise \cite{Elowitz,Rao,Kierzek,Arkin}.

In stochastic models, the underlying Markov process $(X(t))_{t \geq 0}$ is a \emph{pure-jump} process whose state space $\mathcal{S}$ contains all the population size vectors that are \emph{reachable} by the random dynamics. The probability distribution of $(X(t))_{t \geq 0}$ evolves according to a system of linear ordinary differential equations (ODEs), known as the \emph{Chemical Master Equation} (CME) or Forward Kolmogorov Equation \cite{Gillespie:92}. \red{The dimension of the system of ODEs is equal to the number of elements in the state space $\mathcal{S}$, with each element representing a possible combination of reacting species abundances. When $\mathcal{S}$ is finite and small in size, the CME can be solved analytically since it is simply a small and finite system of linear differential equations.
However, for infinite state-spaces an exact solution to the CME is difficult to obtain except in some special cases \cite{Jahnke:07,Grima:12b}.
Beyond these special cases, current methods often rely on truncating the infinite state-space to obtain finite approximations of the CME \cite{FSP}, and then resorting to efficient numerical methods for their solutions. Such methods include Expokit \cite{Sidje:98}, which is based on Krylov Subspace Identification, or the backward Euler method proposed in \cite{Jenkison:12}, among others. Such an approach works well only for relatively small systems, as the curse-of-dimensionality renders the numerical solution of the truncated master equation of larger systems prohibitive. Nevertheless, recent methods based on Tensor Train (TT) and Quantized Tensor Train (QTT) representations \cite{dolgov2012tensor, kazeev2014} show that for CME problems that  admit bounded TT ranks, storage costs and computational complexity that grow linearly in the number of species may be achieved. These and other methods for the numerical solutions of the CME remain active topics of research. }


%

When $\mathcal{S}$ is infinite or very large in size, the most common approach for approximating the solutions of a CME is by simulating a large number of trajectories of the underlying Markov process $(X(t))_{t \geq 0}$, and using the sample values of $X(t)$ to estimate the distribution at time $t$. Such simulations are performed using Monte Carlo procedures such as Gillespie's \emph{stochastic simulation algorithm} (SSA) or its variants \cite{Gillespie:76,Gillespie:77,NR}. Since the simulation time of SSA depends linearly on the number of reactions that occur during the simulation time period, these procedures can be cumbersome for large networks. It is well-known that the stochastic effects caused by the random timing of reactions become less important when the population size is large. The dynamical law of large numbers proved by Kurtz \cite{KurtzLimit} shows that under an appropriate scaling relationship between the population size, reaction rates and the system size, the stochastic model of a reaction network converges to the deterministic model, as the system size goes to infinity. Under this scaling relationship, one can also approximate the stochastic dynamics with certain stochastic differential equations (SDEs) that are easier to simulate and analyze \cite{Kampen,Kurtz2}. However, these SDE approximations can only work when the population sizes of all the species in the reaction network are large, which is often not the case. For a detailed survey on the topic of estimating the solution of a CME, we refer the readers to the paper \cite{GoutsiasSurvey} which contains an exhaustive list of methods for this purpose.

In many biological applications, one in interested in analyzing the long-term behavior or stability properties of a reaction network. This is fairly straightforward for deterministic models because many tools from the theory of ordinary differential equations can be used for this analysis \cite{Khalil}. However, the stability properties of stochastic models for reaction networks are difficult to verify for the following reasons. Let us consider a stochastic reaction network whose dynamics is represented by the Markov process $(X(t))_{t \geq 0}$ with state space $\mathcal{S}$. The evolution of the distributions of this Markov process is given by $(p(t))_{t \geq 0}$ which is the solution of the CME corresponding to the reaction network. Heuristically, we regard the stochastic dynamics to be stable when the family of distributions $(p(t))_{t \geq 0}$ is ``well-behaved" with time. In this paper, we consider several notions of ``well-behaved" dynamics. The strongest of these notions is the concept of \emph{ergodicity} \cite{MeynandTweedieBook} which means that there exists a unique stationary distribution $\pi$ for the Markovian dynamics, such that $p(t) \to \pi$ as $t \to \infty$, irrespective of the initial distribution $p(0)$. This is analogous to having a globally attracting fixed point in the deterministic setting. If $\mathcal{S}$ is finite, the process is ergodic if and only if it is \emph{irreducible}, in the sense that all the states in $\mathcal{S}$ are reachable from each other. It is hence enough to check irreducibility of the process using e.g. matrix methods \cite{Earnshaw,Schnakenberg}. Contrary to this situation, our main interest in this paper is in analyzing the stability properties of stochastic reaction networks with an infinite state space $\mathcal{S}$. Note that in such cases, irreducibility no longer implies ergodicity, since the trajectories of the Markov process may blow up with time (see the carcinogenesis example in the discussion section). In this regard, ergodicity cannot be considered as \emph{a generic property} of reaction networks with infinite state-spaces since both ergodic and non-ergodic processes can be found in nature. Assuming ergodicity without verifying it beforehand seems to be therefore unreasonable from both theoretical and practical perspectives. The direct verification of stability properties like ergodicity is generally not possible as the CME cannot be explicitly solved, except in some restrictive cases \cite{Jahnke:07,Grima:12b}. The common approach of using Monte Carlo simulations for estimating the solutions of a CME is inadequate for assessing the long-term behavior and stability properties of a stochastic reaction network, because one can only simulate finitely many trajectories and those too for a finite amount of time. Some methods for analyzing stability properties without the need for simulations exist, but they either work for specific networks \cite{Grima:12b,Laurenzi:00}, \red{very special classes of networks such as zero-deficiency networks \cite{Anderson}}, or assume system size approximations where the stochastic dynamics is represented by an SDE \cite{Lemarchand,Malek}. Such system size approximations do not hold when some species are present in low copy numbers, and even if they hold, the approximation error generally blows up with time \cite{Kurtz2}. Hence the stochastic dynamics and the corresponding SDE may have completely different long-term behaviors. Our aim, in this paper, is to develop a theoretical and computational framework for analyzing the long-term behavior and stability properties of stochastic models for reaction networks that do not rely on computationally expensive Monte Carlo simulations or on system size approximations of the stochastic dynamics. \red{A similar goal is also achieved in the works \cite{Engblom:12,Rathinam:13} where results on stability and moments bounds are also obtained.} The approach proposed in \cite{Dayar:11} is built upon a Foster-Lyapunov criterion \cite{MeynandTweedieBook} and a quadratic Foster-Lyapunov function in order to estimate the location of the stationary distribution. In the same, yet different, spirit, the proposed approach also relies on a Foster-Lyapunov condition but using a linear Foster-Lyapunov function that allows us to establish ergodicity, moment bounds, moment convergence and the existence of attractive sets for moments. While the approach in \cite{Dayar:11} is fully computational, the one we propose is also theoretical and allows us to conclude on structural properties of classes of networks such as \emph{structural ergodicity}, \emph{structural boundedness of moments} and \emph{structural convergence of moments}. Our approach relies on a mixture of simple ideas from stochastic analysis, linear algebra, polynomial analysis and optimization. \red{Even though our conditions are only sufficient, we demonstrate their broad applicability by successfully establishing stability properties of several reaction networks taken from the literature.}

We mentioned before that the stochastic and the deterministic models of a reaction network are connected through the dynamical law of large numbers \cite{KurtzLimit}. It might be tempting to think that the stability properties of a stochastic model can be assessed by studying the stability properties of the corresponding deterministic model. However in general, the stochastic and deterministic models can have very different stability properties. This is because a deterministic model cannot capture noise induced effects which may have a significant impact on the long-term behavior of a system. For example, in the synthetic Toggle Switch by Gardener \cite{Gardner:00}, the deterministic model exhibits bistability and hence starting from different initial values, the system can converge to two different steady states. On the other hand, the corresponding stochastic model is ergodic (see network \eqref{eq:switch}) and hence the solution of the CME converges to the same stationary distribution irrespective of the initial distribution. A similar phenomenon occurs with the repressilator (see \cite{Elowitz:00} and network \eqref{eq:repressilator}), where the stochastic model is ergodic while the deterministic model exhibits oscillations. On the other hand, it is also possible to find networks for which the deterministic model has a locally asymptotically stable equilibrium point, implying that whenever the initial condition is contained within its region of attraction, the trajectories converge to it. If the initial condition lies outside this region of attraction, then the trajectories of such a network become unbounded with time. In the stochastic setting, the randomness causes each trajectory to leave the region of attraction in finite time, and then become unbounded suggesting that there is no stationary distribution for the dynamics (see network \eqref{eq:networklocal} and Figure \ref{fig:local}). \red{This lack of stationary distribution is because the stochastic dynamics can jump potential wells from one macroscopic fixed point which is stable to another fixed point which is unstable \cite{Vellela:08}. A more striking example of divergent deterministic and stochastic behaviors is given by network \eqref{eq:integrator} (see also Figure \ref{fig:nonergodic}). While the deterministic model has a unique globally stable fixed point, the stochastic model is non-ergodic and all the moments grow unboundedly with time. In this example it is impossible to predict the stochastic behavior from the deterministic model. The above examples illustrate that the stability properties of the stochastic dynamics can, in general, not be assessed from the stability properties of the deterministic dynamics.}

Our results can help in understanding the stability properties of the moments of a Markov process $(X(t))_{t \geq 0}$ representing a reaction network. In particular, we present a method to check if these moments remain bounded with time and if they converge to their steady state values as time goes to infinity. Such results can help in verifying the suitability of a model for a given system and in designing biological controllers that drive the moments to specific steady state values. We provide easily computable bounds for the moments that hold uniformly in time. We also determine bounds for the steady state moment values, which can help in understanding the properties of the steady state distribution, even if this distribution is not explicitly computable. In many biological applications, it is of great interest to explicitly compute the first few moments of the process $(X(t))_{t \geq 0}$ without solving the corresponding CME. One can easily express the dynamics of these moments as a system of ordinary differential equations, but generally this system is not closed when the network has nonlinear interactions. Many moment closure methods that suggest schemes to close these equations to obtain approximations for the moments have been proposed  (see e.g. \cite{Hespanha:08b,Grima:12} and references therein). The results obtained in this paper can be used to ascertain the correctness of a given moment closure method for a specific network (see the example based on the network \eqref{eq:network_CS}). Furthermore, several moment closure methods are developed under an implicit assumption that the \emph{moment-generating} function corresponding to the solution of the CME exists for all times. One of our results provides a way to easily check that this assumption is indeed valid.\\

\textbf{Reaction networks.} Let us now formally describe reaction networks. Motivated by the literature on chemical kinetics, we refer to the network participants as \emph{molecules} which may belong to one of $d$ species $\X{1},\dots,\X{d}$. There are $K$ reactions in the network and for any $k =1,\dots,K$, the \emph{stoichiometric} vector $\zeta_k = (\zeta_{k,1},\dots,\zeta_{k,d})$ denotes the change in the number of molecules in each of the species due to the $k$-th reaction.


\textbf{Deterministic models.} Consider the deterministic model for the reaction network described above. In this setting, the state of the system is described by a vector of \emph{concentrations} of the $d$ species which we denote by $\kappa\in\mathbb{R}_{\ge0}^d$.  The concentration of a species is simply its molecular count divided by the system volume. \red{Let $\tilde{\lambda}_k(\kappa)$ be the \emph{flux} associated with the $k$-th reaction (see \cite{Goutsias}). To ensure positivity of the system, we require that $\tilde{\lambda}_k(\kappa)=0$ whenever $\kappa_i=0$ and $\zeta_{k,i}<0$.} If the initial state is $\kappa_0$, then the evolution of concentrations is given by $(\phi_{\kappa_0}(t) )_{t \geq 0}$ which satisfies the Reaction Rate Equations (RRE) of the form
\begin{align}
\label{deterministicIVP}
\frac{d \phi_{\kappa_0}(t)}{dt} =  \sum_{k=1}^K \tilde{\lambda}_k(\phi_{\kappa_0}(t)) \zeta_k   \ \textnormal{  with } \  \phi_{\kappa_0}(0) = \kappa_0.
\end{align}
We are interested in the long-term behavior and stability of our reaction dynamics. More precisely, we would like to check if the following conditions are satisfied.
\begin{description}
 \item[{\bf DC1}] For any $\kappa_0$, there is a compact set $\mathcal{K}(\kappa_0)$ such that $\phi_{\kappa_0}(t) \in \mathcal{K}(\kappa_0)$ for all $t \geq 0$.
 \item[{\bf DC2}] There exists a compact set $\mathcal{K}_0$ such that for any $\kappa_0$, we have $\phi_{\kappa_0}(t)  \in \mathcal{K}_0$ for large values of $t$.
 \item[{\bf DC3}]  There is a $\kappa_{\textnormal{eq}}$ such that for any $\kappa_0$ we have $\phi_{\kappa_0}(t) \to \kappa_{\textnormal{eq}}$ as $t \to \infty$.
\end{description}
The first condition, {\bf DC1}, says that for any $\kappa_0$, the entire trajectory $(\phi_{\kappa_0}(t) )_{t \geq 0}$ stays within some compact set. We would expect this to be true for most \emph{realistic} systems. Hence a violation of this property may suggest a \emph{flaw} in the deterministic model. The second condition, {\bf DC2}, says that there is an \emph{attractor} set for the dynamics, where all the trajectories eventually lie, irrespective of their starting point. The last condition, {\bf DC3}, says that there is a globally attracting fixed point for the  deterministic model. Using techniques from the theory of dynamical systems \cite{Khalil,Smale:04}, one can verify these conditions, without the need of simulating the deterministic model. There is also a general theory to check condition {\bf DC3} for reaction networks satisfying mass-action kinetics (see \cite{Feinberg1,Feinberg2,HornJackson,Horn1}). Broadly speaking, these three conditions present different ways of saying that the reaction dynamics is ``well-behaved". Our goal in this paper is to develop a theoretical and computational framework for verifying conditions similar to ${\bf DC1}, {\bf DC2}$ and $ {\bf DC3}$ for stochastic models of reaction networks.


\textbf{Stochastic models.} Consider the stochastic model corresponding to the reaction network described above. In this setting, the firing of reactions are discrete events and the state of the system refers to the vector of molecular counts of the $d$ species. When the state is $x$, the $k$-th reaction fires after a random time which is exponentially distributed with rate $\lambda_k(x)$. The functions $\lambda_1,\dots,\lambda_K$ are known as the \emph{propensity} functions in the literature. \red{To ensure positivity of the system, we require that if $x+\zeta_k\notin\mathbb{N}_0^d$, then $\lambda_k(x)=0$, where $\N_0$ is the set of non-negative integers.} The dynamics can be represented by the Markov process $(X_{x_0}(t))_{ t \geq 0}$ where $x_0$ is the initial state. Note that if $X_{x_0}(t) = (X_1(t),\dots,X_d(t))$, then $X_i(t)$ is the \emph{number} of molecules of $\X{i}$ at time $t$.

It is important to select a suitable state space $\mathcal{S}$ for the Markov process representing the reaction dynamics. We choose $\mathcal{S}$ to be a non-empty subset of $\N^d_0$ satisfying the following properties:
\begin{itemize}
\item[(A)] If $x \in \mathcal{S}$ and $\lambda_k(x) > 0$ for some $k=1,\dots,K$, then $x + \zeta_k \in \mathcal{S}$.
\item[(B)] There is no proper subset $\mathcal{S}_1 \subset \mathcal{S}$ satisfying part (A).
\end{itemize}
Observe that part (A) ensures that if $x_0 \in \mathcal{S}$ then $X_{x_0}(t) \in \mathcal{S}$ for all $t \geq 0$ and hence $\mathcal{S}$ can be taken to be the state space of all the Markov processes describing the stochastic reaction network with an initial state $x_0$ in $\mathcal{S}$. Part (B) implies that the reaction dynamics cannot be contained in a proper subset of $\mathcal{S}$. The role of this assumption will become clear in the next section, when we discuss the issue of state space \emph{irreducibility}. Note that in certain cases, such as the \emph{pure-birth} network $\phib  \longrightharpoonup \X{1}$, a suitable state space satisfying the above criteria cannot be found. There also exist cases where the above criteria restricts the choice of state space. For example, for the \emph{pure-death} network $  \X{1}\longrightharpoonup \phib$, the only possible choice for state space is $\mathcal{S} =\{0\}$. Finally we remark that if the reactions in a network satisfy a conservation relation then the state space must be chosen with an initial condition in mind. For example, for the network $\X{1} \rightleftharpoons \X{2}$, the sum of molecular counts of $\X{1} $ and $\X{2}$ is preserved by the reactions. Hence if we wish to study the stochastic dynamics with the initial sum as $n$, then the correct choice for state space is $ \mathcal{S} =\{(x_1,x_2) \in \N^2_0 : x_1+x_2 = n\}$.


Let $\mathcal{P}( \mathcal{S} )$ denote the space of probability distributions over $\mathcal{S}$, endowed with the weak topology \red{which is metrized by the Prohorov metric} (see \cite{EK}).
For any $x,y \in \mathcal{S}$ let $p_x(t,y)$ denote the following probability
\begin{align}
\label{transition_probabilities}
p_x(t,y) = \Prob\left( X_x(t) = y \right).
\end{align}
Defining $p_x(t)(A) = \sum_{y \in A} p_x(t,y)$, for any $A \subset \mathcal{S}$, we can view $p_x(t)$ as an element in $\mathcal{P}(\mathcal{S})$. In fact, $p_x(t)$ is the
distribution at time $t$ of the Markov process $(X_x(t))_{t \geq 0}$.
The dynamics of $p_x(t)$ is given by the Chemical Master Equation (CME) which has the following form:
\begin{align}
\label{cme}
\frac{ d p_{x}(t,y) }{dt} = \sum_{k=1}^K \left( p_{x}(t, y -\zeta_k) \lambda_k(y - \zeta_k) - p_{x}(t, y )  \lambda_k(y) \right),
\end{align}
where $p_x(0,y) = 1$ if $x =y$ and $p_{x}(0,y) = 0$ for all $y \neq x$.
Theoretically, one can find $p_{x}(t,y)$ for any $t \geq 0$ and $y \in \mathcal{S}$, by solving this system. However this system consists of as many equations as the number of elements in $\mathcal{S}$. Hence an explicit solution is only possible when $\mathcal{S}$ is finite, which only happens in very restrictive cases where all the reactions preserve some conservation relation. Typically, $\mathcal{S}$ is infinite and solving this system analytically or even numerically is nearly impossible, except in some restrictive cases (see \cite{Jahnke:07}). From now on, we assume that $\mathcal{S}$ is infinite.%

The above discussion shows that at the level of distributions, we can view the stochastic dynamics $(X_{x_0}(t))_{t \geq 0}$ as the deterministic dynamics $(p_{x_0}(t))_{ t \geq 0 }$, which satisfies the CME.
However, the major difficulty in analyzing this deterministic dynamics is that it occurs over an infinite dimensional space $\mathcal{P}(\mathcal{S})$. Nevertheless we can recast the conditions {\bf DC1}, {\bf DC2} and {\bf DC3} in the stochastic setting as below.
\begin{description}
 \item[{\bf SC1}] For any $x_0$, there is a compact set $\mathcal{K}(x_0) \subset \mathcal{P}(\mathcal{S})$ such that $p_{x_0}(t)\in \mathcal{K}(x_0)$ for all $t \geq 0$.
 \item[{\bf SC2}] There exists a compact set $\mathcal{K}_0 \subset \mathcal{P}(\mathcal{S})$ such that for any $x_0 \in \mathcal{S}$ we have $p_{x_0}(t) \in \mathcal{K}_0$ for large values of $t$.
 \item[{\bf SC3}] There is a $\pi \in \mathcal{P}(\mathcal{S})$ such that for any $x_0$ we have $p_{x_0}(t) \to \pi$ as $t \to \infty$.
\end{description}
Each of the above conditions give an important insight about the long-term behavior and stability of the stochastic dynamics. The first condition, {\bf SC1}, says that for every $\epsilon \in(0,1)$ we can find a finite set $A_\epsilon \subset \mathcal{S}$ such that each $p_{x_0}(t)$ puts at least $(1-\epsilon)$ of its mass in $A_\epsilon$. \red{In other words, the probability that the state of the underlying Markov process at any time $t$ is inside $A_\epsilon $ is greater than $(1-\epsilon)$.} We would expect this to be true for most realistic models.
If condition {\bf SC2} holds then the evolution of distributions have a compact attractor set in $\mathcal{P}(\mathcal{S})$, where all the trajectories eventually lie irrespective of their starting point. This suggests that in the long run, the family of processes $\{ (X_{x_0}(t))_{t \geq 0} : x_0 \in \mathcal{S} \}$, spend most of their time on the same set of states. The last condition {\bf SC3} says that the evolution of distributions have a globally attracting fixed point $\pi$. If this holds, then the Markov process representing the reaction dynamics is \emph{ergodic} with $\pi$ as the unique stationary distribution. For understanding the long-term behavior of a stochastic process, ergodicity is a desirable property to have. In the long-run, the proportion of time spent by any trajectory of an ergodic process, in any subset of the state space is equal to the stationary probability of that subset (see \eqref{resultLLN}).  In other words, information about the stationary distribution can be obtained by observing just one trajectory for a sufficiently long time.
Such a result can have important applications. For example, consider a culture with a large number of identical cells with each cell having the same reaction network. If we can show that this intracellular network is ergodic, then by observing the long-term reaction dynamics in a single cell, using for example. time-lapse microscopy, we can obtain statistical information about all the cells at stationarity. Conversely, ergodicity allows us to obtain the stationary distribution of a single-cell by observing the distribution over the population, using for example flow cytometry.

In this paper we develop a general framework for checking conditions {\bf SC1}, {\bf SC2} and {\bf SC3}. However, the scope of our paper is broader than that. As mentioned in the introduction, we obtain easily computable bounds for the statistical moments of the underlying Markov process and investigate when these moments converge with time. We also present conditions for the distribution of the process to be \emph{light-tailed}.

\section*{Results}

\subsection*{Preliminaries}

In this section we discuss the main results of our paper. In particular, we explain how conditions {\bf SC1}, {\bf SC2} and {\bf SC3} can be verified without having to simulate the trajectories of the Markov process representing the reaction dynamics. Intuitively, these conditions can only hold if the Markov process has a low probability of hitting states that have a very large \emph{size}. In our case, the states are vectors in $\R^d$ and so we can measure their \emph{size} by using any norm on $\R^d$. The central theme of this paper is to demonstrate that for many networks, long-term behavior can be easily analyzed by choosing the \emph{right} norm for measuring the state sizes. This \emph{right} norm has the form
\begin{align}
\label{defn_vnorm}
\|x\|_v = \sum_{i=1}^d v_i |x_i|,
\end{align}
where $v$ is a positive vector in $\R^d$ satisfying the following condition.
\begin{condition}[Drift-Diffusivity Condition]
\label{MainCondition}
For a positive vector $v \in \R^d$, \red{there exist positive constants $c_1,c_2,c_3,c_4$ and a nonnegative constant $c_5$} such that for all $x \in \mathcal{S}$
\begin{subequations}\label{reqd_conditions}
  \begin{align}
   & \sum_{k =1 }^K \lambda_k(x) \langle  v, \zeta_k \rangle  \leq c_1 - c_2 \langle  v , x \rangle  \textnormal{ and }\label{eq:cond1} \\
    & \sum_{k =1 }^K \lambda_k(x) \langle  v, \zeta_k \rangle^2  \leq c_3 + c_4 \langle  v , x \rangle + c_5 \langle v,x \rangle^2.\label{eq:cond2}
  \end{align}
\end{subequations}
\end{condition}
Here $\langle \cdot, \cdot \rangle$ denotes the standard inner product on $\R^d$. If we consider the process $(  \|X_{x_0}(t)\|_v )_{t \geq 0}$, then its dynamics can be seen to have two components \emph{drift} and \emph{diffusion} which have the form $\sum_{k =1 }^K \lambda_k(x) \langle  v, \zeta_k \rangle$ and $\sum_{k =1 }^K \lambda_k(x) \langle  v, \zeta_k \rangle^2$ respectively when $X_{x_0}(t) = x$ (see page 2 in the Supplementary Material \textbf{S1}). Condition \ref{MainCondition} gives upper-bounds for the magnitude of these two components and hence we call it the \emph{drift-diffusivity condition} (abbreviated to Condition {\bf DD} from now on; the abbreviations {\bf DD1} and {\bf DD2} stand for the first and second inequality, respectively). Observe that when the process $(  \|X_{x_0}(t)\|_v )_{t \geq 0}$ goes above $c_1/c_2$ then it experiences a negative drift, suggesting that it will move downwards. This fact will be crucial for our analysis.

For now, we assume that a vector $v$ satisfying Condition {\bf DD} has been found. In later sections we demonstrate how $v$ can be determined for a large class of networks by solving suitably constructed optimization problems.

For any positive integer $r$, let $m^{r}_{x_0}(t)$ denote the $r$-th moment of $\|X_{x_0}(t)\|_v$ defined by
\begin{align}
\label{defn_mr}
m^{r}_{x_0}(t) = \E \left(  \|X_{x_0}(t)\|^r_v \right) = \sum_{y \in \mathcal{S}} \|y\|^r_v p_{x_0}(t,y).
\end{align}
Similarly let $\Psi^r(x_0,t)$ denote the $r$-th moment of $X_{x_0}(t)$ at time $t$. Then $\Psi^r(x_0,t)$ is a tensor of rank $r$ whose entry at index $(i_1,\dots, i_r )\in \{1,2,\dots,d\}^r$ is given by
\begin{align}
\label{momentdefnition}
\Psi^r_{i_1\dots i_r}(x_0,t) = \sum_{y \in \mathcal{S} } y_{i_1} \dots y_{i_r} p_{x_0}(t,y),
\end{align}
where $y =(y_1,\dots,y_d)$ and $p_{x_0}(t)$ is the distribution of $X_{x_0}(t)$.

Suppose that for some positive constants $r$ and $C_r(x_0)$ we have
\begin{align}
\label{unif_moment_bd_r}
\sup_{t \geq 0} m^r_{x_0}(t) \leq  C_r(x_0).
\end{align}
For any $M > 0$, let $K_M$ be the compact (finite) set defined by $K_M = \{ x \in \mathcal{S}  : \|x\|_v \leq M  \}$ and let $K^{c}_M$ denote its complement. Markov's inequality (see \cite{Kal}) implies that for any $\epsilon >0$ we can choose $M$ large enough to satisfy
\begin{align*}
\sup_{t \geq 0} p_{x_0}(t,K^{c}_M) = \sup_{t \geq 0}  \Prob \left(  \| X_{x_0}(t)\|^{r}_v > M^r \right) \leq \sup_{t \geq 0} \E \left(  \|X_{x_0}(t)\|^r_v \right)  \leq \frac{C_r(x_0)}{M^r} < \epsilon.
\end{align*}
Hence Prohorov's theorem (see Chapter 3 in \cite{EK}) ensures that condition {\bf SC1} holds. Similarly we can prove that condition {\bf SC2} will hold if for some $r >0$ there exists a constant $\hat{C}_r$ such that
\begin{align}
\label{lim_moment_bd_r}
\limsup_{t \to \infty} m^r_{x_0}(t) \leq \hat{C}_r \textnormal{ for all } x_0 \in \mathcal{S}.
\end{align}
Relations \eqref{unif_moment_bd_r} and \eqref{lim_moment_bd_r} give uniform and asymptotic upper-bounds for $m^{r}_{x_0}(t) $.
Using these relations we can also obtain uniform and asymptotic upper-bounds for the entries of $\Psi^r(x_0,t)$.
Such moment bound results have applications in queuing theory and control theory (see \cite{Pemantle}). In Theorem \ref{moment_boundsthm} we show that under certain conditions, \eqref{unif_moment_bd_r} and \eqref{lim_moment_bd_r} hold and the upper-bounds can be easily computed.


Instead of the $r$-th moment of the process $( \| X_{x_0}(t) \|_v )_{t \geq 0}$, one can ask if the exponential moment of this process is uniformly bounded from above. This will happen if for some $\gamma > 0$ we have
\begin{align}
\label{uniform_exp_bound}
\sup_{t \geq 0} \E \left( e^{ \gamma \|X_{x_0}(t)\|_v } \right) = \sup_{t \geq 0} \sum_{y \in \mathcal{S}} e^{ \gamma \|y\|_v} p_{x_0}(t,y) <  \infty.
\end{align}
If \eqref{uniform_exp_bound} holds, then the distribution $p_{x_0}(t)$ is \emph{light-tailed} (a distribution is called light-tailed if its tails are majorized by an exponential decay) uniformly in $t$. This shows that all the cumulants of the distribution $p_{x_0}(t)$ exist, which is an important result for the following reason. There is a considerable body of research dedicated to estimating the moments of the process $(X_{x_0}(t))_{t \geq 0}$ directly without computing the distribution functions $p_{x_0}(t)$. For any integer $r>0$, one can easily write the differential equations for the dynamics of the first $r$ moments.
However when the reaction network has nonlinear interactions, this system of equations is not closed for any $r$. Various moment closure methods (see \cite{MomentClosure,Pendar:13}) exist that specify ways to close these equations artificially and estimate the moments approximately. A popular moment closure method is the cumulant-neglect method which ignores the higher order cumulants of the distribution $p_{x_0}(t)$ for all $t \geq 0$. Of course this method is only valid when the higher order cumulants exist. This is guaranteed if \eqref{uniform_exp_bound} holds. In Theorem \ref{unif_exp_moment_bounds} we give conditions for verifying \eqref{uniform_exp_bound}.

We now come to the question of checking condition {\bf SC3} which says that the process $(X_{x_0}(t))_{t \geq 0}$ is ergodic. This can only happen if the state space $\mathcal{S}$ is \emph{irreducible}, which means that all the states are accessible from each other. Recall the definition of $p_x(t,y)$ from \eqref{transition_probabilities}. Mathematically, we say that $\mathcal{S}$ is irreducible if for all $x,y \in \mathcal{S}$, we have $p_x(t_1,y) > 0$ and $p_y(t_2,x) > 0$ for some $t_1,t_2 > 0$.
In order to check the irreducibility of $\mathcal{S}$, one has to verify that there is no proper subset $\mathcal{S}_1 \subset \mathcal{S}$, such that once the process reaches a state in $\mathcal{S}_1$, it stays in $\mathcal{S}_1$ forever. 
For reaction networks with mass-action kinetics, methods for checking irreducibility have recently been reported in \cite{Pauleve:13}  and \cite{Gupta:13}. These methods can be extended to situations where the propensity functions are positive in the positive orthant. When the propensity functions vanish inside the positive orthant, the problem of checking irreducibility can become much more complicated, and to the best of our knowledge no methods exist in the literature for this purpose.

We mentioned before that the vector $v$ is chosen so that the process $(  \|X_{x_0}(t) \|_v )_{t \geq 0}$ has a negative drift at large values.
Assuming irreducibility, this is sufficient to verify ergodicity of $(X_{x_0}(t))_{t \geq 0}$ (see Proposition \ref{ergodicproposition}).

Suppose that condition {\bf SC3} is satisfied and the process $(X_{x_0}(t))_{t \geq 0}$ is ergodic with stationary distribution $\pi$. For any positive integer $r$, let $\Pi^r$ denote the $r$-th moment of the stationary distribution $\pi$. Then $\Pi^r$ is a tensor of rank $r$ defined in the same way as $\Psi^r(x_0,t)$ (see \eqref{momentdefnition}), with $p_{x_0}(t,y)$ replaced by $\pi(y)$.
Using Theorem \ref{moment_boundsthm} we can determine the values of $r$ for which $\Pi^r$ is finite (componentwise) and $\Psi^r(x_0,t) \to \Pi^r$ as $t \to \infty$ (see Theorem \ref{whichmomentsconverge}). We can also identify functions $f : \mathcal{S} \to \R$ for which
\begin{align}
\label{ergodicconvergenceoff}
\lim_{t \to \infty} \E( f(X_{x_0}(t)) ) = \sum_{y \in  \mathcal{S} } f(y) \pi(y) < \infty
\end{align}
holds for any $x_0 \in \mathcal{S}$. If $f$ is such a function, then the ergodic theorem for Markov processes (see \cite{Norris}) says that
\begin{align}
\label{resultLLN}
\lim_{t \to \infty} \frac{1}{t} \int_{0}^{t} f(X_{x_0}(s))ds = \sum_{y \in \mathcal{S}} f(y) \pi(y) \textnormal{ almost surely},
\end{align}
for any $x_0 \in \mathcal{S}$.
Lastly, we also obtain conditions to check if the stationary distribution $\pi$ is \emph{light-tailed} (see Theorem \ref{stationary_exp_moment_bounds}).

\subsection*{General Results}

In this section, we formally present the main results of our paper. Their proofs are given in the Supplementary Material \textbf{S1}.

\textbf{Moment bounds.} Our first result establishes that for certain values of $r$, we can obtain uniform and asymptotic moment bounds for the $r$-th moment of the process $( \|X_{x_0}(t)\|_v)_{t \geq 0}$.
\begin{theorem}
\label{moment_boundsthm}
\red{Assume that Condition {\bf DD} holds.}
Let $r_{\textnormal{max}}$ be given by
\begin{align}
\label{defn_rmax}
r_{\textnormal{max}} = \left\{
\begin{array}{cc}
1 + \frac{2c_2}{c_5} & \textnormal{if} \ c_5 > 0 \\
\infty & \textnormal{if} \ c_5 = 0.
\end{array} \right.
\end{align}
For any positive integer $r$, if $r < r_{ \textnormal{max} }$ then there exist positive constants $C_r(x_0)$ and $\hat{C}_r$ such that \eqref{unif_moment_bd_r} and \eqref{lim_moment_bd_r} hold.
\end{theorem}
The values of the constants $C_r(x_0)$ and $\hat{C}_r$ can be explicitly computed using a recursive relationship (see the Supplementary Material \textbf{S1}). Note that if $v= (v_1,\dots,v_d)$, then for any $y = (y_1,\dots,y_d) \in \mathcal{S}$ we have $y_i\leq \|y\|_v / v_i$ for any $i$.
Hence for any $i_1,\dots,i_r \in \{1,2,\dots,d\}$ we have $\Psi^r_{i_1\dots i_r}(x_0,t) \leq m^{r}_{x_0}(t) / \prod_{j=1}^r  v_{i_j}  $
Therefore using Theorem \ref{moment_boundsthm}, we can obtain uniform and asymptotic moment bounds for the reaction dynamics $(X_{x_0}(t))_{t \geq 0}$ (see the Supplementary Material \textbf{S1}).

Observe that if $c_5 = 0$ then $r_{\textnormal{max}} = \infty$. In this case, Theorem \ref{moment_boundsthm} says that for each positive integer $r$ and $x_0 \in \mathcal{S}$ there exists a constant $C_r(x_0)$ such that \eqref{unif_moment_bd_r} holds. By showing that we have a $C >0$ such that $C_r(x_0) \leq r! C^r$ for all positive integers $r$, we obtain our next result, which gives sufficient conditions to check \eqref{uniform_exp_bound}.
\begin{theorem}[Uniform Light-Tailedness]
\label{unif_exp_moment_bounds}
Suppose that  Condition {\bf DD} holds with $c_5 = 0$. \red{Given an initial state $x_0 \in \mathcal{S}$} there exists a $\gamma > 0$ such that
\begin{align*}
\sup_{t \geq 0} \E \left( e^{ \gamma \|X_{x_0}(t)\|_v} \right) = \sup_{t \geq 0} \sum_{y \in \mathcal{S}} e^{ \gamma \|y\|_v} p_{x_0}(t,y) <  \infty.
\end{align*}
\end{theorem}

\textbf{Ergodicity and Moment Convergence.} The next result verifies the ergodicity of a reaction network satisfying  Condition {\bf DD}. It follows from Theorem 7.1 in Meyn and Tweedie \cite{Meyn1}.
\begin{proposition}{\bf (Ergodicity)}
\label{ergodicproposition}
Assume that the state space $\mathcal{S}$ of the Markov process $(X_{x_0}(t))_{t \geq 0}$ is irreducible \red{and Condition {\bf DD1} holds.}
Then this process is exponentially ergodic in the sense that there exists a unique distribution $\pi \in \mathcal{P}(\mathcal{S})$ along with constants $B,c >0$ such that for any $x_0 \in \mathcal{S}$
\begin{align*}
 \sup_{A \subset \mathcal{S}} \left| p_{x_0}(t,A) - \pi(A) \right| \leq B e^{-c t} \textnormal{ for all } t\geq 0.
\end{align*}
 \end{proposition}
This result says that as $t \to \infty$, the distribution $p_{x_0}(t)$ converges to $\pi$ exponentially fast. Henceforth we assume that the process $(X_{x_0}(t))_{t \geq 0}$ is ergodic with stationary distribution $\pi$.

Let $f : \mathcal{S} \to \R$ be a function such that for some positive integer $r < ( r_{ \textnormal{max} } - 1)$, there exists a $C> 0$ satisfying
$|f(x)| \leq C(1 + \|x\|_v^r)$ for all $x \in \mathcal{S}$. Using Theorem \ref{moment_boundsthm} we can prove that for such a $f$, the relations \eqref{ergodicconvergenceoff} and \eqref{resultLLN} hold. As a consequence we obtain the following result about the convergence of moments with time.
\begin{theorem}[Moment Convergence]
\label{whichmomentsconverge}
\red{Assume that Condition {\bf DD} holds.}
Let $r$ be any positive integer satisfying $r < ( r_{ \textnormal{max} } - 1)$. Then $\Pi^r$ is finite (componentwise) and $\Psi^r(x_0,t) \rightarrow
\Pi^r$ as $t \to \infty$.
\end{theorem}
If $f(x) = \|x\|^r_v$ then Theorem \ref{moment_boundsthm}  and \eqref{ergodicconvergenceoff} imply that for any positive integer $r< (r_{ \textnormal{max}}-1)$ there exists a positive constant $\hat{C}_r$ such that
 \begin{align}
 \label{momentboundonstationarydistr}
\sum_{y \in \mathcal{S}} \| y\|^r_v \pi(y) \leq \hat{C}_r.
\end{align}
In particular, if $c_5 =0$ then $r_{ \textnormal{max} } = \infty$ and \eqref{momentboundonstationarydistr} holds for each $r$.
By proving the existence of a constant $C >0 $ such that $\hat{C}_r \leq r! C^r$ for all positive integers $r$ we get our last result which shows that the stationary distribution is \emph{light-tailed}.
\begin{theorem}[Light-Tailedness at stationarity]
\label{stationary_exp_moment_bounds}
Suppose that  Condition {\bf DD} holds with $c_5 = 0$. Then there exists a $\gamma > 0$ such that
\begin{align*}
\sum_{y \in \mathcal{S}} e^{ \gamma \|y\|_v} \pi(y) <  \infty.
\end{align*}
\end{theorem}

The framework described above is very general and can be applied to any network that satisfies  Condition {\bf DD}. In what follows, we specialize the results for two wide classes of networks with mass-action kinetics, namely reaction networks with monomolecular and bimolecular reactions. It will be, however, pointed out in the examples that the scope \red{of our approach} is much broader since more general propensities, \red{such as those involving Hill functions or more general mass-action kinetics, can be considered.}

\section*{Methods}

\red{Using the analytical tools developed in the previous sections, several general results can be stated for the class of \red{unimolecular reaction networks} and \red{bimolecular reaction networks}. In what follows, when we say that a moment is bounded, we mean that it is bounded uniformly in time (as in \eqref{unif_moment_bd_r}). This can be established using Theorem \ref{moment_boundsthm} once Condition \textbf{DD} is verified. Furthermore, when we say that a moment is globally converging, we mean that it converges to its equilibrium value as time tends to infinity, irrespective of the initial state $x_0$. Once, Condition \textbf{DD} is verified, this can established using Theorem \ref{whichmomentsconverge}.}

\red{The main aim of the section is to develop a theoretical and computational framework for checking Condition \textbf{DD}.}

\subsection*{\red{Results for stochastic unimolecular reaction networks}}

Let us then consider \red{a unimolecular reaction network }which involves $d$ species that interact through $K$ reaction channels of the form: 
\begin{equation}\label{eq:affine_r}
  \begin{array}{lcllcllcl}
    \phib&\stackrel{k_0^i}{\longrightharpoonup}&\X{i},& \X{i}&\stackrel{k_i^0}{\longrightharpoonup}&\phib,&  \X{i}&\stackrel{k_{i}^{\ell}}{\longrightharpoonup}&\sum_{j=1}^d\nu_i^{j\ell}\X{j}
  \end{array}
\end{equation}
\red{where $i=1,\ldots,d$, $\ell\in\{1,\ldots,N_i\}$, $N_i>0$ and $\nu_i^{j\ell}\in\mathbb{N}_0$. }The reaction rates $k_0^i$, $k_i^0$ and $k_{i}^{\ell}$ are positive real numbers. In accordance with \eqref{cme}, the reactions are indexed from $n=1$ to $K$, and corresponding propensities and stoichiometries are denoted by $\lambda_n(x)$ and $\zeta_n$, respectively.

\textbf{Motivations.} The \red{unimolecular} case may seem quite restrictive at first sight and not of particular practical interest. We demonstrate below that, on the contrary, the proposed results on unimolecular reaction networks complete existing ones and are, therefore, of practical and theoretical interests. Although some explicit solutions for the CME are indeed known for some particular unimolecular reactions \cite{Jahnke:07}, it is still unknown whether the CME admits an closed-form solution for all possible type of unimolecular reactions. Note that we assume here that no simplification nor assumption is made on the problem, we are dealing with the very general unimolecular case.

The results developed of this section are useful in several ways. First of all, all types of unimolecular reactions can be handled with the proposed approach, making it more general than existing ones in this regard. Moreover, given a specific reaction network, the method allows one to establish whether a unique stationary distribution exists \emph{without solving the CME}. This is particularly important since unimolecular networks \emph{may not be ergodic}. \red{In this case, the network can exhibit unstable behaviour which may suggest a flaw in the model if the considered real-world system exhibits stable trajectories. Moreover, in certain design applications such as those in synthetic biology, it seems natural to design networks that have well-behaved dynamics. Checking ergodicity provides a convenient way to determine if the network dynamics is well-behaved.} Note, furthermore, that  it is, in general, difficult to infer ergodicity directly from the solution of the CME (when it is known) since proving the existence of a unique globally attractive stationary distribution amounts to check convergence of the solution to the CME to the same distribution for all possible initial distributions, which are in infinite number in our setup. This fact is even more true when large networks are considered since the explicit form of the solution to the CME is, in this case, very intricate \cite{Jahnke:07}. The proposed results allow \red{one} to circumvent this difficulty and demonstrate that ergodicity can be assessed by very simple means, i.e. using basic notions of linear algebra. The results can be furthermore used to assess the \emph{structural ergodicity} of a reaction network, that is, the ergodicity of a network for any combination of the rate parameters, by very simple means. This very strong and practically relevant notion is extremely difficult, again, to check from the solution of the CME since it would require to check the convergence of the solution of the CME to the same stationary distribution for all initial conditions and all positive values of the rate parameters, a very cumbersome task, even for small networks. Finally, the results pertaining on \red{unimolecular networks} will also turn out to play an important role in the ergodicity analysis of \red{bimolecular reaction networks}.


%


\textbf{Theoretical results.} Let us start with several theoretical results that characterize the long-term behavior of \red{unimolecular networks} of the form \eqref{eq:affine_r}.
\begin{proposition}[Ergodicity of \red{unimolecular} networks]\label{th:nomerg}
Let us consider the general \red{unimolecular} reaction network \eqref{eq:affine_r} and assume that the state-space of the underlying Markov process is irreducible. Let the matrices $A\in\mathbb{R}^{d\times d}$ and $b\in\mathbb{R}^d_{\ge0}$, \red{$||b||\ne0$}, be further defined as
  \begin{equation}\label{eq:dksodjoglab1}
  \sum_{n=1}^K\lambda_n(x)\langle v,\zeta_n\rangle=x^\T Av+b^\T v.
  \end{equation}
  Then, the following statements are equivalent:
  \begin{enumerate}
    \item The matrix $A$ is Hurwitz-stable, i.e. all its eigenvalues lie in the open left half-plane.
    \item There exists a vector $v\in\mathbb{R}_{>0}^d$ such that $Av<0$.
  \end{enumerate}
  Moreover, when one of the above statements holds, the Markov process describing the reaction network is exponentially ergodic and all the moments are bounded and globally converging.\mendex
\end{proposition}
The above result shows that, for \red{unimolecular networks}, ergodicity and the existence of moment bounds can be directly inferred from the properties of the matrix $A$ defined in \eqref{eq:dksodjoglab1}. The second statement, which characterizes the Hurwitz-stability of $A$ in an implicit way, will turn out to play a key role in the analysis of \red{unimolecular and bimolecular reaction networks} since checking whether $Av<0$ for some $v>0$ is a linear programming problem.

\red{It is important to stress that, in the result above, if we simply demand that the moments be bounded and converging, then $A$ may be allowed to have zero eigenvalues in certain cases. Note, however, that the moments will converge to values that may depend on the initial conditions.}

In the case that the structure of the network (the reactions and the stoichiometries) is exactly known, but that the reaction rates are subject to uncertainties, the above theorem can be \emph{robustified} to account for these uncertainties. To this aim, suppose that the matrix $A$ depends on a vector $\delta \in [-1,1]^\eta$ where $\eta\in\mathbb{N}$ is the number of distinct uncertain parameters. We write this matrix as $A(\delta)$ and assume that there exists a matrix $A_+ \in \R^{d \times d}$ satisfying the following properties:
\begin{enumerate}
\item $A(\delta) \leq A_+$ (in the componentwise sense) for all $\delta\in[-1,1]^\eta$
\item There exists a $\delta^*\in[-1,1]^\eta$ such that $A_+ = A(\delta^*)$.
\end{enumerate}
Note that such a matrix $A_+$ may not exist, especially when some entries are not independent. However, when $A_+$ exists we have the following result.
  \begin{proposition}[Robust ergodicity]
   Let us consider the general \red{unimolecular reaction network} \eqref{eq:affine_r} described by some uncertain matrices $A(\delta)$ and $b(\delta)$, $||b(\delta)||\ne0$. Assume further the matrix $A(\delta)$ admits the upper-bound  $A_+$ defined above and that the state-space of the underlying Markov process is irreducible for all uncertain parameter values $\delta\in[-1,1]^\eta$. Then, the following statements are equivalent:
  \begin{enumerate}
    \item The matrix $A(\delta)$ is Hurwitz-stable for all $\delta\in[-1,1]^\eta$.
    \item The matrix $A_+$ is Hurwitz-stable.
    \item There exists a positive vector $v\in\mathbb{R}^d$ such that $A_+v<0$.
  \end{enumerate}
  Moreover, when one of the above statements holds, the Markov process describing the reaction network is robustly exponentially ergodic and all the moments are bounded and globally converging.\mendex
\end{proposition}

\red{Observe that checking the Hurwitz-stability property of each $A(\delta)$ is equivalent to checking it for only $A_+$. Hence we can conclude that, in this case,  \emph{checking ergodicity of a family of networks is not more complicated than checking ergodicity of  a single network}.} The case when the matrix $A_+$ is not defined is more complicated and is discussed in the supplementary material \textbf{S1}.


\textbf{Computational results.} We now present several computational results that accompany the theoretical results of the previous section. \red{ It is possible to extract many computational results from our general framework, but for simplicity we only address the problems of checking ergodicity and computing the first-order moment bounds.} The asymptotic first-order moment bound, defined in Theorem \ref{moment_boundsthm}, is given by $\widehat{C}_1=c_1/c_2$. So the question arises: what is the smallest value for such a ratio? Or, in other words, what is the smallest attractive compact set for the first-order moment of $\langle v,X(t)\rangle$? Several numerical methods, solving exactly or approximately this problem, are discussed in the supplementary material \textbf{S1}. One of them is the following optimization problem which is \emph{fully equivalent to Proposition \ref{th:nomerg}}:
\begin{opt}
Let us consider the general \red{unimolecular reaction network} \eqref{eq:affine_r} and assume that the state-space of the underlying Markov process is irreducible. Assume further that the optimization problem
\begin{equation}\label{eq:optlf1}
\begin{array}{rcl}
   \max_{z,v}z& \textnormal{s.t.}&z>0,v>\eps\\
   && (zI+A)v\le0
 \end{array}
 \end{equation}
  is feasible with $(z^*,v^*)$ as minimizer. Then, we have $\widehat{C}_1^*\le b^\T v^*/z^*$ and Proposition \ref{th:nomerg} holds.
\end{opt}
%

A striking feature about the above optimization program is that the numbers of variables and constraints are given by $d+1$ and $2d+1$, respectively. This means that the optimization problem scales \emph{linearly} with respect to the number of species ($d$) in the network, and is independent of the number of reactions $K$. Therefore, from the point of view of this optimization problem, the size of a \red{unimolecular network} can be identified with the number of species, and not the number of reactions. The above optimization problem can be efficiently solved using a bisection algorithm over $z$ that is globally and geometrically converging to $z^*$. Each iteration consists of solving a linear program, a class of optimization problems known to be very tractable, and for which numerous advanced solvers exist \cite{Boyd:04}. These properties, altogether, make the overall approach highly scalable, which is necessary for dealing with very large networks.

\subsection*{Results for stochastic bimolecular reaction networks}

Similar results are now presented for stochastic \emph{bimolecular reaction networks} which, in addition to the \red{unimolecular  reactions} \eqref{eq:affine_r}, also involve \red{bimolecular reactions}
 of the form:
\begin{equation}\label{eq:quad_r}
    \begin{array}{lcllcl}
    \X{i}+\X{j}&\stackrel{k_{ij}^{\ell}}{\longrightharpoonup}& \sum_{m=1}^d\nu_{ij}^{m\ell}\X{m},&   \X{i}+\X{j}&\stackrel{k_{ij}^{0}}{\longrightharpoonup}&\phib
  \end{array}
\end{equation}
\red{where $i,j=1,\ldots,d$, $\ell\in\{1,\ldots,N_{ij}\}$, $N_{ij}>0$, and $\nu_{ij}^{m\ell}\in\mathbb{N}_0$}. The reaction rates $k_{ij}^{\ell}$ and $k_{ij}^{0}$ are positive real numbers.


\textbf{Theoretical results for \emph{bimolecular  networks}.} When \emph{bimolecular reaction networks} of the form \eqref{eq:affine_r}-\eqref{eq:quad_r} are considered, the left-hand side of condition \eqref{eq:cond1} can be expressed as
\begin{equation}\label{eq:quadgen}
  \sum_{i=1}^K\lambda_k(x)\langle v,\zeta_k\rangle=x^\T M(v)x+x^\T Av+b^\T v
\end{equation}
where $M(v)\in\mathbb{R}^{d\times d}$ is symmetric, $A\in\mathbb{R}^{d\times d}$ and $b\in\mathbb{R}_{\ge0}^{d}$. Let $S:=\begin{bmatrix}
  \zeta_1 & \ldots & \zeta_K
\end{bmatrix}$ be the stoichiometry matrix of the \emph{bimolecular reaction network} \eqref{eq:affine_r}-\eqref{eq:quad_r}, and let $S_q$ be the restriction of $S$ to \emph{bimolecular reactions}, only. Further define a set
\begin{equation*}
  \mathcal{N}_q:=\left\{v\in\mathbb{R}^d:\ v>0,\ v^\T S_q=0\right\}.
\end{equation*}
When $v\in\mathcal{N}_q$, the quadratic term $x^\T M(v)x$ in \eqref{eq:quadgen} vanishes, and equation \eqref{eq:quadgen} reduces to
\begin{equation*}
  \sum_{i=1}^K\lambda_k(x)\langle v,\zeta_k\rangle=x^\T Av+b^\T v
\end{equation*}
which is exactly the same expression as in the case of \red{unimolecular}  networks. This means that, with the additional constraint that $v\in\mathcal{N}_q$, all the results derived for \red{unimolecular} networks directly apply to \emph{bimolecular networks} as well. This allows us to obtain the following result.
\begin{proposition}[Ergodicity of \emph{bimolecular networks}]\label{th:nomerg_q}
Let us consider the \emph{bimolecular reaction network} of the form \eqref{eq:affine_r}-\eqref{eq:quad_r} such that $||b||\ne0$ in \eqref{eq:quadgen} and assume that the state-space of the underlying Markov process is irreducible. Assume further that the network admits a non-empty $\mathcal{N}_q$.

If there exists a vector $v\in\mathcal{N}_q$ such that the inequality $Av<0$ holds, then the stochastic \emph{bimolecular reaction network} \eqref{eq:affine_r}-\eqref{eq:quad_r} is ergodic and all the moments are bounded and globally converging.\mendex
\end{proposition}

It  is important to mention that the existence of a non-empty set $\mathcal{N}_q$ is a prerequisite for utilizing the above result. \red{Non-emptiness of $\mathcal{N}_q$ is equivalent to the existence of a conservation relation for all the bimolecular reactions, i.e. the value of (at least) a positive linear combination of the species populations remains unchanged when any of the bimolecular reactions fires. Note that this definition extends to more general mass-action kinetics as well. A necessary condition for the non-emptiness of $\mathcal{N}_q$ is that $S_q$ is not full-row rank.} This non-emptiness condition may seem restrictive at first sight, but it will be shown that several important reaction networks from the literature satisfy this condition.

Whenever $\mathcal{N}_q$ is empty or there is no $v\in\mathcal{N}_q$ such that $Av<0$ holds, the next result can be used.
\begin{proposition}[Ergodicity of \red{bimolecular  networks}]\label{th:nomerg_q2}
Let us consider the \red{bimolecular reaction network} of the form \eqref{eq:affine_r}-\eqref{eq:quad_r}  such that $||b||\ne0$ in \eqref{eq:quadgen}  and assume that the state-space of the underlying Markov process is irreducible. Assume further that one of the following statements holds:
\begin{enumerate}
  \item There exists $v\in\mathbb{R}_{>0}^d$ such that $Av<0$ and $M(v)\le0$ hold.
  \item There exists $v\in\mathbb{R}_{>0}^d$ such that $M(v)$ is negative definite.
\end{enumerate}

Then, the stochastic \red{bimolecular reaction network} \eqref{eq:affine_r}-\eqref{eq:quad_r} is ergodic and all the moments up to order $(\lfloor1+2c_2/c_5\rfloor-2)$ are bounded and globally converging.\mendex
\end{proposition}
In the above result, the first statement can be checked using a linear program since the inequalities are componentwise. Checking the second statement, however, requires a semidefinite program, which is a more general convex program, that can be solved using solvers such as SeDuMi \cite{Sturm:01a} and SDPT3 \cite{Tutuncu:03}. More details on the above result can be found in the supplementary material \textbf{S1}.

\textbf{Computational results for \red{bimolecular networks}.} It is shown here that, once again, the theoretical results can be easily turned into linear programs that can be checked in a very efficient way. The following result is the numerical translation of Proposition \ref{th:nomerg_q}.
\begin{opt}
Let us consider a \red{bimolecular reaction network} \eqref{eq:affine_r}-\eqref{eq:quad_r} and assume that the state-space of the underlying Markov process is irreducible. Assume further that $\mathcal{N}_q\ne\emptyset$ and that the optimization problem
\begin{equation}\label{eq:optlf3}
\begin{array}{rcl}
   \max_{z,v}z&  \textnormal{s.t.} &z>0,v>\eps\\
   && (zI+A)v\le0\\
   &&v^\T S_q=0.
 \end{array}
 \end{equation}
 is feasible with $(z^*,v^*)$ as minimizer. Then, we have $\widehat{C}_1^*\le b^\T v^*/z^*$
  and Proposition \ref{th:nomerg_q} holds.
\end{opt}
%
The computational complexity of this optimization problem scales linearly with the number of species and can therefore be solved for large networks. 

The following optimization problem is the computational counterpart of the first statement of Proposition \ref{th:nomerg_q2}.
\begin{opt}
Let us consider a \red{bimolecular reaction network} of the form \eqref{eq:affine_r}-\eqref{eq:quad_r} and assume that the state-space of the underlying Markov process is irreducible. Assume further that the nonlinear optimization problem
\begin{equation}\label{eq:optlf4}
\begin{array}{rcl}
   \max_{z,v}z&\textnormal{s.t.} &z>0, v>\eps\\
   && (zI+A)v\le0\\
   &&M(v)\le0.
 \end{array}
 \end{equation}
 is feasible with $(z^*,v^*)$ as minimizer. Then, we have $\widehat{C}_1^*\le b^\T v^*/z^*$
  and Proposition \ref{th:nomerg_q2} holds.
\end{opt}
The above optimization problem does not scale as nicely as \eqref{eq:optlf3} since, in the worst case, the number of constraints related to $M(v)$ is quadratic in the number of species. 
The problem, however, remains tractable due to the linear programming structure.


\subsection*{Qualitative differences between deterministic and stochastic dynamics}

\red{ In this section we illustrate that stochastic and deterministic models of the same reaction network may exhibit very different qualitative behaviors. Therefore assessing ergodicity or the convergence of moments of a stochastic model from the stability properties of the corresponding deterministic model is, in general, incorrect. To support this claim, we consider two reaction networks.}

\red{\subsubsection*{Jumping potential wells}}

\red{ Our first example shows that stochastic dynamics can jump potential wells and leave the stability regions of the deterministic dynamics, resulting in an unstable behavior. Consider the following reaction network:
\begin{equation}\label{eq:networklocal}
        \begin{array}{rcl}
          \phib &\stackrel{\alpha\beta}{\longrightharpoonup}& \Xo\\
           \Xo&\stackrel{\alpha+\beta}{\longrightharpoonup}&\phib\\
          \Xo+\Xo&\stackrel{1}{\longrightharpoonup}&3\Xo
        \end{array}
      \end{equation}
      where $0<\alpha<\beta$. The deterministic dynamics for this network is given by
      \begin{equation}\label{eq:deterlocal}
        \dot{\kappa}=f(\kappa):=\kappa^2-(\alpha+\beta)\kappa+\alpha\beta
      \end{equation}
      where $\kappa\in\mathbb{R}_{\ge0}$ denotes the concentration of $\mathbf{S}$. The fixed points for the dynamics are $\kappa_-=\alpha$ and $\kappa_+=\beta$, respectively. From the graph $\{(\kappa,f(\kappa))\in\mathbb{R}_{\ge0}\times\mathbb{R}:\ \kappa\in\mathbb{R}_{\ge0}\}$, it is immediate that the fixed point $\kappa_-=\alpha$ is locally asymptotically stable with the region of attraction as $[0,\beta)$ while the other fixed point $\kappa_+=\beta$ is unstable.}

\red{We now consider the stochastic version of this network and let $\mathbb{A}$ be the generator of the corresponding Markov process. For the identity function $f(x) =x$ we have
      \begin{equation}\label{eq:djjsdjkjfdsl}
        \mathbb{A}f(x)=\frac{1}{2}x^2-\left(\alpha+\beta+\frac{1}{2} \right) x+\alpha\beta
      \end{equation}
 The polynomial on the right-hand side has two positive roots that are
      \begin{equation}
            x_\pm=\alpha+\beta+\frac{1}{2}\pm\sqrt{ \left(\alpha+\beta+\frac{1}{2}\right)^2-2\alpha\beta}.
      \end{equation}
      This means that for all $x\in\mathbb{N}_0$ satisfying $x\ge 1+x_+$, we have $\mathbb{A}f(x)\ge\eps$, for some $\eps>0$, implying that the drift is positive. So if the state of the state of the network reaches a value that is greater than $1+x_+$, then there is a possibility that the trajectories become unbounded with time.}

      \red{To demonstrate this, we pick $\alpha=7/2$ and $\beta=21/2$. In such a case, the largest root of the polynomial on the right-hand side of \eqref{eq:djjsdjkjfdsl} is $x_+=\left(29+\sqrt{547}\right)/2\simeq26.194>\beta$. We can see that the region where the drift $\mathbb{A}f(x)$ is negative is actually larger than the region of attraction of the locally asymptotically stable fixed point for the deterministic dynamics. This is due to the fact that the propensity function of the \red{bimolecular reaction} differs from whether we are in the deterministic or in the stochastic setting.}

\begin{figure}[!ht]
    \centering
 \includegraphics[width=0.75\textwidth]{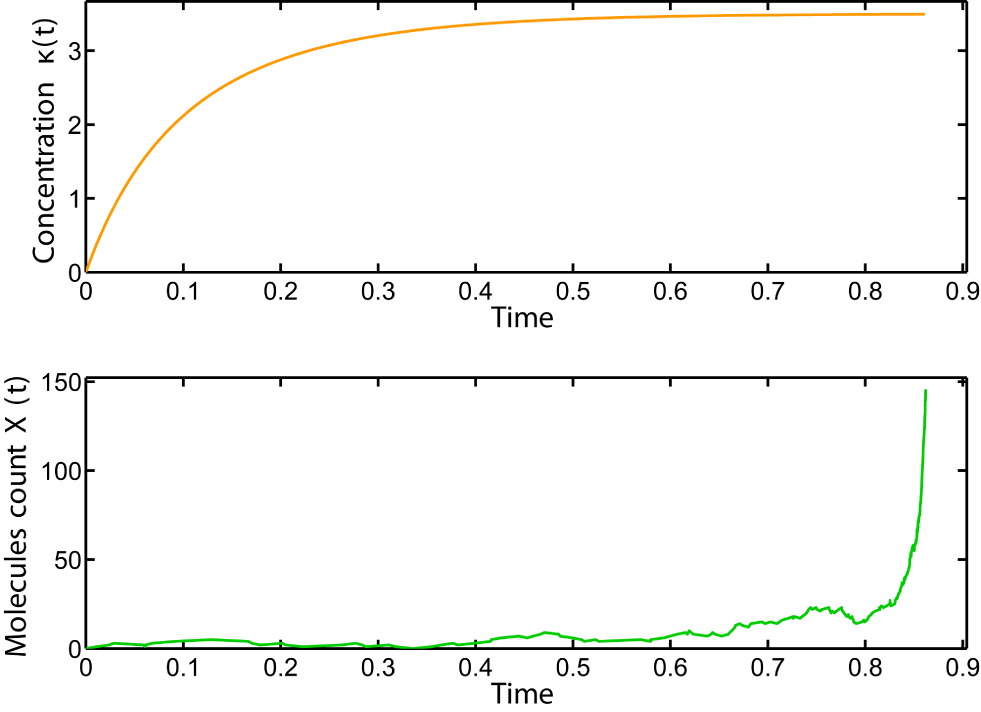}
  \caption{\textbf{Trajectory of the state of the deterministic system \eqref{eq:deterlocal} with initial condition $\kappa_0=0$ (top); Sample path of the Markov process describing the network \eqref{eq:networklocal} with initial condition $x_0=0$ (bottom).} Whereas the trajectory of the state of the deterministic model converges to a stationary value, the trajectory of the state of the stochastic model goes unbounded.} \label{fig:local}
\end{figure}

       \red{Let us now set the initial condition $\kappa_0=0$ for the deterministic model and  $x_0=0$ for the stochastic one. Note that they both lie within the region of attraction of the fixed point of the deterministic dynamics and in the region of negative drift for the stochastic dynamics. We then perform 1000 SSA runs over 100 seconds and stop the simulation when the propensity function $x(x-1)/2$ of the \red{bimolecular reaction} exceeds the value corresponding to $15000$ molecules (approx. $1.12\times10^{8}$). At this rate value, the \red{bimolecular reaction} fires, on average, every $10^{-8}$ seconds, leading to an explosion of the state of the system and to unbounded trajectories. Out of 1000 SSA runs, all were stopped before the end of the simulation time-period (100 seconds). This behavior strongly indicates that the system is \emph{not ergodic} despite the the fact that the deterministic model has a locally asymptotically stable fixed point. Figure \ref{fig:local} illustrates the above discussion.}

\red{\subsubsection*{Globally stable deterministic dynamics does not imply moments stability}}

      \red{ In the previous example, the stochastic and deterministic behaviors were different, but one can still understand stochastic instability through the deterministic model. The deterministic dynamics posseses a region in which the solutions explode and the randomness in the stochastic dynamics allows it to enter this region in finite time and grow unbounded thereafter. We now present an example which is more striking in the sense that the deterministic model cannot be used in any way to infer the instability of the stochastic model. In this example, the deterministic dynamics has a unique fixed point which is exponentially stable, while the stochastic dynamics is not ergodic with all its moments growing unboundedly with time.}

      \red{Consider the reaction network given by
     \begin{equation}\label{eq:integrator}
        \begin{array}{rcl}
          \phib &\stackrel{1}{\longrightharpoonup}& \X{1}\\
          \phib &\stackrel{1}{\longrightharpoonup}& \X{2}\\
          \X{1}+\X{2}&\stackrel{1}{\longrightharpoonup}&\phib.
        \end{array}
      \end{equation}}
 \red{ Let $\kappa\in\mathbb{R}^2_{\ge0}$ be the vector of concentrations. The state of the deterministic model evolves according to
   \begin{equation}\label{eq:detdskdl}
     \begin{array}{lcl}
        \dot{\kappa}_1(t)&=&1-\kappa_1(t)\kappa_2(t)\\
        \dot{\kappa}_2(t)&=&1-\kappa_1(t)\kappa_2(t).
     \end{array}
   \end{equation}
   Assume that the initial conditions satisfy $\kappa_2(0)-\kappa_1(0)=\alpha$, for some $\alpha\in\mathbb{R}$. Then we have the following result.}
    \red{
    \begin{theorem}
      The unique equilibrium point of the dynamics \eqref{eq:detdskdl} given by
       \begin{equation}
     \kappa_1^*=\dfrac{1}{2}\left(-\alpha+\sqrt{\alpha^2+4}\right)\ \textnormal{and}\  \kappa_2^*=\dfrac{1}{2}\left(\alpha+\sqrt{\alpha^2+4}\right).
   \end{equation}
   is globally exponentially stable.
    \end{theorem} }

\red{In the stochastic setting, the picture is completely different as the next result indicates.}
%
\red{
\begin{theorem}
The Markov process corresponding to the stochastic model of network \eqref{eq:integrator} is not ergodic and all its moments grow unboundedly with time. Moreover, if $X_1(0)-X_2(0)=\alpha$ for some $\alpha>0$, we have that $\E[X_1(t)-X_2(t)]=\alpha$ for all $t\ge0$.
\end{theorem}}

\begin{figure}[!ht]
\centering
\includegraphics[width=0.75\textwidth]{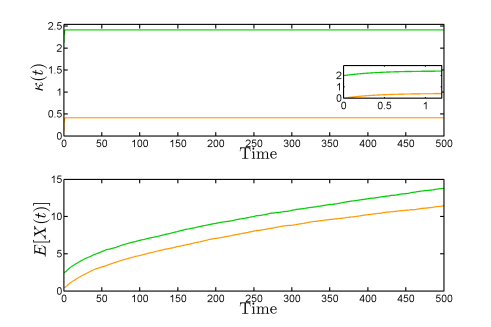}
  \caption{\textbf{Comparison of the trajectories of the deterministic and stochastic (first-order moments) models of the reaction network \eqref{eq:integrator} with initial condition $\kappa_1(0)=0$, $\kappa_2(0)=2$, $X_1(0)=0$ and $X_2(0)=2$ for the deterministic (top) and stochastic dynamics (bottom), respectively.} We can see that while the deterministic trajectories converge to their equilibrium point, the first-order moments grow without bound.}\label{fig:nonergodic}
\end{figure}

\red{To illustrate this result, we simulate the deterministic and the stochastic process (10000 SSA runs) for $\kappa_1(0)=0$, $\kappa_1(0)=\alpha$, $X_1(0)=0$, $X_2(0)=\alpha$ and  $\alpha=2$. The results are shown in Figure \ref{fig:nonergodic}.}

\subsection*{Finding an attractive compact set for the first-order moments}
\begin{figure}[!ht]
    \centering
   \includegraphics[width=0.75\textwidth]{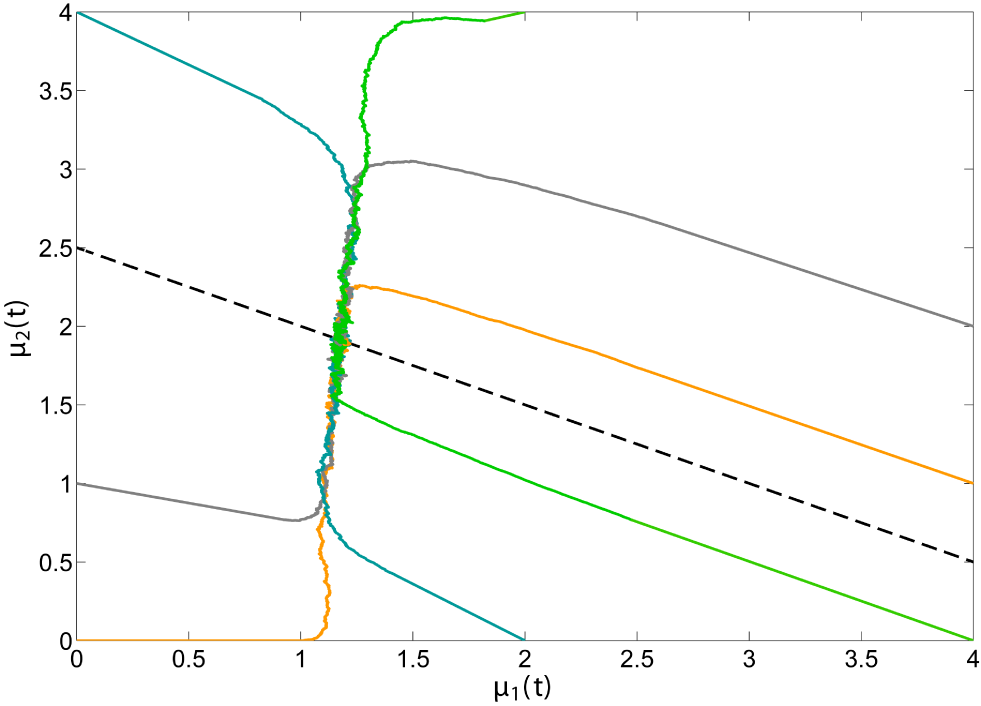}
  \caption{\textbf{Trajectories of the first order moments {\small$\mu_1(t)=\mathbb{E}[X_1(t)]$} and $\mu_2(t)=\mathbb{E}[X_2(t)]$ of network \eqref{eq:network_CS} for different initial conditions (averaging is performed over 5000 cells).} The trajectories converge to the unique steady-state value located inside the compact set (the surface below the dashed line), very close to the boundary.}\label{fig:compact_set}
\end{figure}

The goal of this section is to compute a compact set that is attractive for the first-order moment of $\langle v,X(t)\rangle$ using the optimization problems \eqref{eq:optlf1} or \eqref{eq:optlf3}. Due to the moment closure problem \cite{MomentClosure}, analytical expressions for the steady-state values of the moments of \red{bimolecular reaction networks} are not available, and hence this is an important class of networks to analyze. Consider the following \red{bimolecular reaction network}
 \begin{equation}\label{eq:network_CS}
  \begin{array}{rclcrcl}
    \phib &\stackrel{k}{\longrightharpoonup}&\X{1},&&   \X{1}&\stackrel{\gamma_1}{\longrightharpoonup}&\phib \\
    \X{1}+\X{1}&\stackrel{k_{12}}{\longrightharpoonup}&\X{2},&&   \X{2}&\stackrel{k_{21}}{\longrightharpoonup}&\X{1}+\X{1}\\
    \X{2}&\stackrel{\gamma_2}{\longrightharpoonup}&\phib .
  \end{array}
\end{equation}
representing a dimerization process, i.e. $\X{1}$ dimerizes to $\X{2}$. It is easily seen that this network is irreducible since any point in the state-space can be reached from any other point in a finite number of reactions having nonzero propensities. Choosing $v$ in $\mathcal{N}_q$, e.g. $v^\T=\begin{bmatrix}
   1 & 2
 \end{bmatrix}$, yields that $c_1^*=k$ and $c_2^*=\min\{\gamma_1,\gamma_2\}$, hence the network is exponentially ergodic, and all the moments are bounded and converging. On solving the optimization problem \eqref{eq:optlf3} with numerical values $k=1$, $\gamma_1=\gamma_2=0.2$, $k_{12}=1$ and $k_{21}=0.1$, we get that $\hat{C}_1  = c_1^*/c_2^*=5$ which coincides with the theoretical value \red{$k/\min\{\gamma_1,\gamma_2\}$}. One can regard $ \{(x_1,x_2) \in \mathbb{R}_{>0}^2 : v^\T x \leq \hat{C}_1\}$ to be an attractive compact set in which the first-order moments of $\langle v,X(t) \rangle$ eventually lie. To validate this calculation, Monte-Carlo simulations were performed which yield
\begin{equation}
  \lim_{t\to\infty}\E[\langle v,X(t) \rangle ]=5.024\pm0.05,
\end{equation}
showing the correctness of the attractive compact set. 
To further illustrate this result, several trajectories of $\E[X_1(t)]$ and $\E[X_2(t)]$ for different initial conditions are plotted in Figure~\ref{fig:compact_set}.

We now discuss how the computation of an attractive compact set for the first-order moments can be used to assess whether a closure method leads to a result that is consistent with the stochastic dynamics. The idea is to check whether the closed system converges towards a value which lies within the compact set. Let us consider the reaction network \eqref{eq:network_CS} and close the first-order moments equations by neglecting the second order cumulant, i.e. neglecting the variance. By doing so, we get the model
\begin{equation}
  \begin{array}{lcl}
    \dot{\tilde{\mu}}_1(t)&=&k-\gamma_1\tilde{\mu}_1(t)-k_{12}\tilde{\mu}_1(t)(\tilde{\mu}_1(t)-1)+2k_{21}\tilde{\mu}_2(t)\\
    \dot{\tilde{\mu}}_2(t)&=&k_{12}\tilde{\mu}_1(t)(\tilde{\mu}_1(t)-1)-\gamma_2\tilde{\mu}_2(t)
  \end{array}
\end{equation}
where $\tilde{\mu}_1$ and $\tilde{\mu}_2$ are the approximate first-order moments of the system.  The unique positive equilibrium point for this model is given by
\begin{equation}
  \begin{array}{lcl}
    \tilde{\mu}_1^*&=&\dfrac{1}{2k}\left(-\gamma_1+\dfrac{k_{12}\gamma_2}{\gamma_2+k_{21}}+\sqrt{\Delta}\right)\\
    \tilde{\mu}_2^*&=&\dfrac{k_{12}}{2(\gamma_2+k_{21})}\tilde{\mu}_1^*(\tilde{\mu}_1^*-1)
  \end{array}
\end{equation}
where $\Delta=\left(-\gamma_1+\dfrac{k_{12}\gamma_2}{\gamma_2+k_{21}}\right)^2+\dfrac{4kk_{12}\gamma_2}{\gamma_2+k_{21}}$.

With the same parameter values as before, we find that $\tilde{\mu}_1^*=1.6238$ and $\tilde{\mu}_2^*=1.6881$ and therefore $v^\T\tilde{\mu}^*=5$ for $v^\T=\begin{bmatrix}
  1 & 2
\end{bmatrix}$, showing that the state of the closed system converges to the
boundary of the compact set. Note that SSA also predicts that the trajectories of the first-order moments converge to the boundary of this set.
However the actual equilibrium values for the first-order moments of the stochastic dynamics are $\mu_1^*\simeq 1.1450$ and $\mu_2^*\simeq1.9350$, which differ from the ones obtained with the closure method. This discrepancy is expected since the variance has been neglected.

This example shows how attractive compact sets for the moments can be used as a test for the momet-closure methods by checking whether the closed system predicts trajectories that that converge inside those compact sets. However, note that in the current state, these compact sets can only be used to obtain a lower bound on the closure-error whenever the trajectories of the closed dynamics converge to a point outside the compact set. In such a case, the lower bound on the closure-error $\eps$ is simply given by the distance between the equilibrium point of the closed-system
 \begin{equation}
   \eps\ge\inf_{\theta\in C}||\tilde{\mu}^*-\theta||_2
 \end{equation}
 where $C$ is the attractive (convex) compact set and $\tilde{\mu}^*$ is the equilibrium point  of the closed dynamics.


\subsection*{Feedback loop}
\begin{figure}[!ht]
    \centering
    \includegraphics[width=0.45\textwidth]{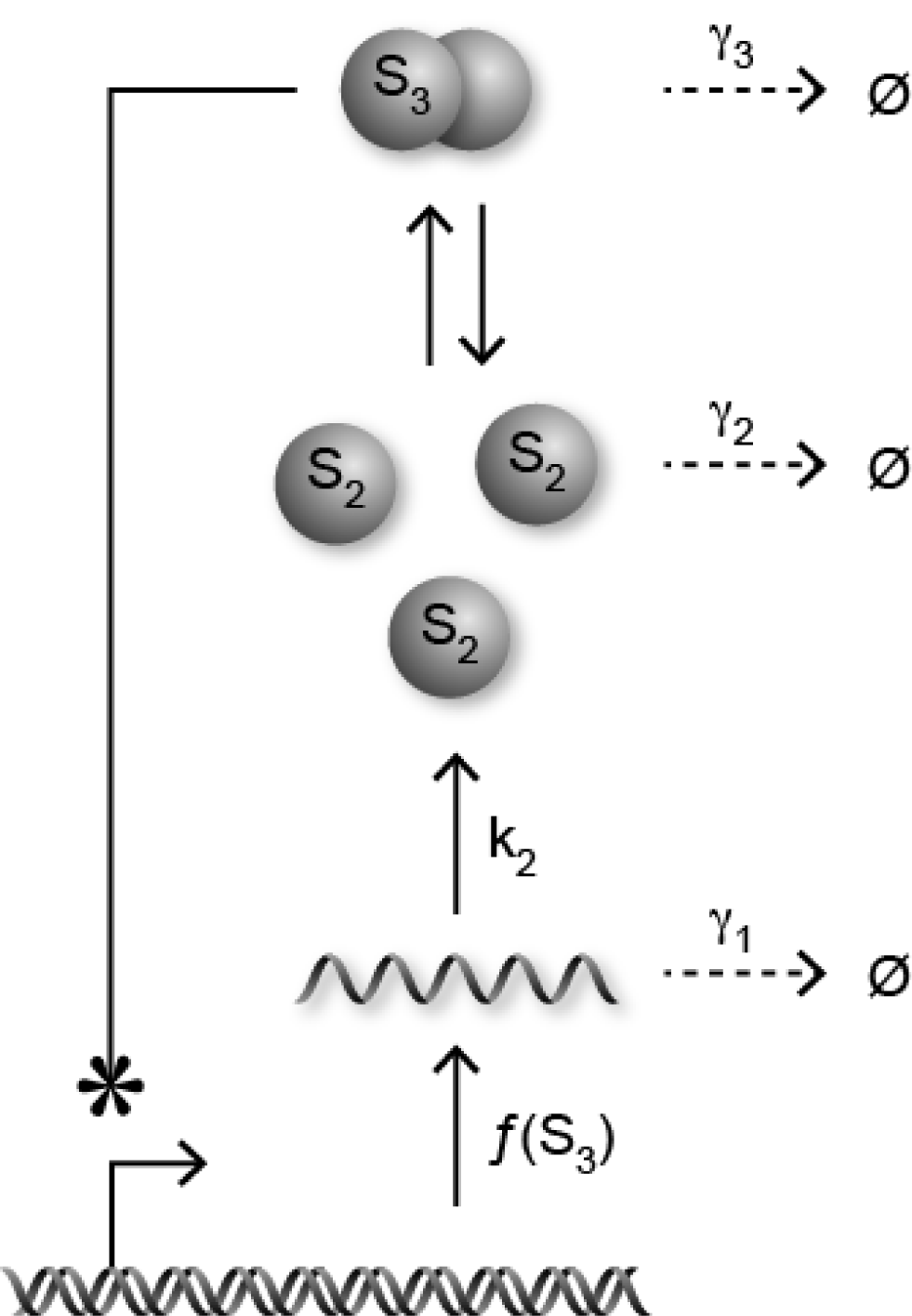}
  \caption{\textbf{Feedback loop with arbitrary feedback rule.}}\label{fig:feedback}
\end{figure}

Let us consider the feedback loop network of Figure~\ref{fig:feedback} represented by the reaction network
\begin{equation}\label{eq:feedback_dimer}
  \begin{array}{rclcrcl}
  \X{1}&\stackrel{k_2}{\llongrightharpoonup}&\X{1}+\X{2},&&  \phib &\stackrel{f(\X{3})}{\llongrightharpoonup}&\X{1}\\
    \X{3}&\stackrel{k_{32}}{\llongrightharpoonup}&\X{2}+\X{2},&&  \X{2}+\X{2}&\stackrel{k_{23}}{\llongrightharpoonup}&\X{3}\\
    \X{i}&\stackrel{\gamma_i}{\llongrightharpoonup}&\phib .
  \end{array}
\end{equation}
where $\X{1}$ is mRNA and $\X{2}$ is the corresponding protein. The dimer  $\X{3}$ acts back on the gene expression through an arbitrary bounded nonnegative function $f(\cdot)$.

%
 We have the following result:
\begin{result}\label{res:feedback_dimer}
  For any positive values of the rate parameters and any bounded nonnegative function $f(\cdot)$, the feedback loop with dimerization \eqref{eq:feedback_dimer} is ergodic and all the moments are bounded and globally converging.
\end{result}


\subsection*{Stochastic switch}
Let us consider the stochastic switch of \cite{Tian:06} described by the \red{unimolecular} stochastic reaction network
\begin{equation}\label{eq:switch}
  \begin{array}{lclclcl}
    \phib &\stackrel{f_1(\boldsymbol{S_2^1})}{\longrightharpoonup}&\boldsymbol{S_1^0},&&  \boldsymbol{S_1^0}&\stackrel{k_1}{\longrightharpoonup}&\boldsymbol{S_1^0}+\boldsymbol{S_1^1}\\
    \phib &\stackrel{f_2(\boldsymbol{S_1^1})}{\longrightharpoonup}&\boldsymbol{S_2^0},&&  \boldsymbol{S_2^0}&\stackrel{k_2}{\longrightharpoonup}&\boldsymbol{S_2^0}+\boldsymbol{S_2^1}\\
     \boldsymbol{S_i^j}&\stackrel{\gamma_{i,j}}{\longrightharpoonup}&\phib .
  \end{array}
\end{equation}
Above $\boldsymbol{S_i^0}$ and $\boldsymbol{S_i^1}$ represent mRNAs and proteins of gene $i$, respectively. The functions $f_1(\cdot)$ and $f_2(\cdot)$ are arbitrary bounded nonnegative functions. 
 We have the following result:
\begin{result}
  For any positive values of the rate parameters and any bounded nonnegative functions $f_1(\cdot)$ and $f_2(\cdot)$, the stochastic switch \eqref{eq:switch}  is ergodic and all the moments are bounded and globally converging.
\end{result}

\subsection*{Repressilator}
\begin{figure}[!ht]
    \centering
    \includegraphics[width=0.45\textwidth]{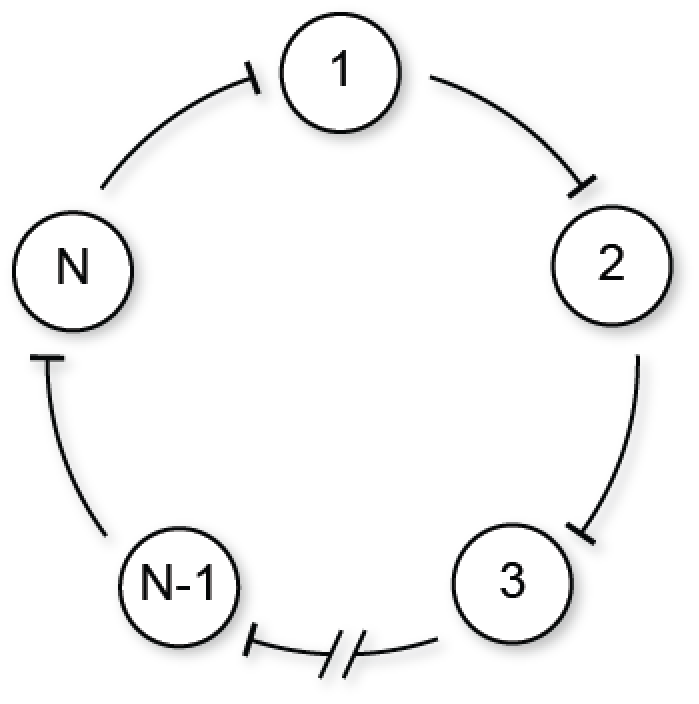}
  \begin{flushleft}
    \caption{\textbf{$N$-gene repressilator.}}\label{fig:repressilator}
  \end{flushleft}
\end{figure}

We consider here the stochastic repressilator of Figure~\ref{fig:repressilator} (see also \cite{Elowitz:00}) involving $N$ genes.

The reaction network corresponding to this $N$-gene repressilator is given by
\begin{equation}\label{eq:repressilator}
  \begin{array}{lcl}
    \phib&\stackrel{f_1(\boldsymbol{S^1_N})}{\longrightharpoonup}&\boldsymbol{S^1_1}\\
    \phib&\stackrel{f_2(\boldsymbol{S^1_1})}{\longrightharpoonup}&\boldsymbol{S_2^1}\\
    \phib&\stackrel{f_3(\boldsymbol{S_2^1})}{\longrightharpoonup}&\boldsymbol{S_3^1}\\
    \vdots&\vdots & \vdots\\
    \phib&\stackrel{f_N(\boldsymbol{S^1_{N-1}})}{\longrightharpoonup}&\boldsymbol{S^1_N}\\
     \boldsymbol{S^1_1}&\stackrel{k_1}{\longrightharpoonup}&\boldsymbol{S^1_1}+\boldsymbol{S^2_1}\\
     \boldsymbol{S^1_2}&\stackrel{k_2}{\longrightharpoonup}&\boldsymbol{S^1_2}+\boldsymbol{S^2_2}\\
     \boldsymbol{S^1_3}&\stackrel{k_3}{\longrightharpoonup}&\boldsymbol{S^1_3}+\boldsymbol{S^2_3}\\
    \vdots&\vdots & \vdots\\
    \boldsymbol{S^1_N}&\stackrel{k_n}{\longrightharpoonup}&\boldsymbol{S^1_N}+\boldsymbol{S^2_N}\\
     \boldsymbol{S^1_i}&\stackrel{\gamma_i}{\longrightharpoonup}&\phib,\ i=1,\ldots,N\\
     \boldsymbol{S^2_i}&\stackrel{\delta_i}{\longrightharpoonup}&\phib,\ i=1,\ldots,N
  \end{array}
\end{equation}
where $f_i(x)=\alpha_i+\beta_i/(1+x^n)$, $\alpha_i,\beta_i,n>0$. Above, $\boldsymbol{S^1_i}$ and $\boldsymbol{S^2_i}$ are the mRNA and protein corresponding to gene $i$. We have the following result:
\begin{result}
For any positive values of the rate parameters $k_i,\gamma_i,\delta_i,\alpha_i,\beta_i$ and $n$, the stochastic $N$-gene repressilator \eqref{eq:repressilator} is ergodic and all the moments are bounded and globally converging.
\end{result}

\subsection*{Stochastic SIR model}
We consider here the following SIR-model, similar to the one in \cite{Chen:05}, defined as
\begin{equation}\label{eq:SIR}
  \begin{array}{rclcrclcrcl}
    \phib &\stackrel{k_s}{\longrightharpoonup}&\boldsymbol{S},&&\phib &\stackrel{k_i}{\longrightharpoonup}&\boldsymbol{I},   &&   \boldsymbol{S}&\stackrel{\gamma_s}{\longrightharpoonup}&\phib \\
    \boldsymbol{I}&\stackrel{\gamma_i}{\longrightharpoonup}&\phib ,  &&  \boldsymbol{R}&\stackrel{\gamma_r}{\longrightharpoonup}&\phib , && \boldsymbol{S}+\boldsymbol{I}&\stackrel{k_{si}}{\longrightharpoonup}&2\boldsymbol{I}\\
    \boldsymbol{I}&\stackrel{k_{ir}}{\longrightharpoonup}&\boldsymbol{R},&&  \boldsymbol{R}&\stackrel{k_{rs}}{\longrightharpoonup}&\boldsymbol{S}.
    \end{array}
\end{equation}
where birth and death reactions represent people entering and leaving the process, respectively. The only \red{bimolecular reaction} is the contamination reaction which turns one susceptible person into an infectious one. The two last reactions represent how infectious people are recovering and how recovered people become susceptible again. We then have the following result:
\begin{result}
  For any positive values of the rate parameters, the SIR-model \eqref{eq:SIR} is ergodic and all the moments are bounded and globally converging.
\end{result}

\subsection*{Circadian clock}
\begin{figure}[!ht]
\centering
  \includegraphics[width=0.55\textwidth]{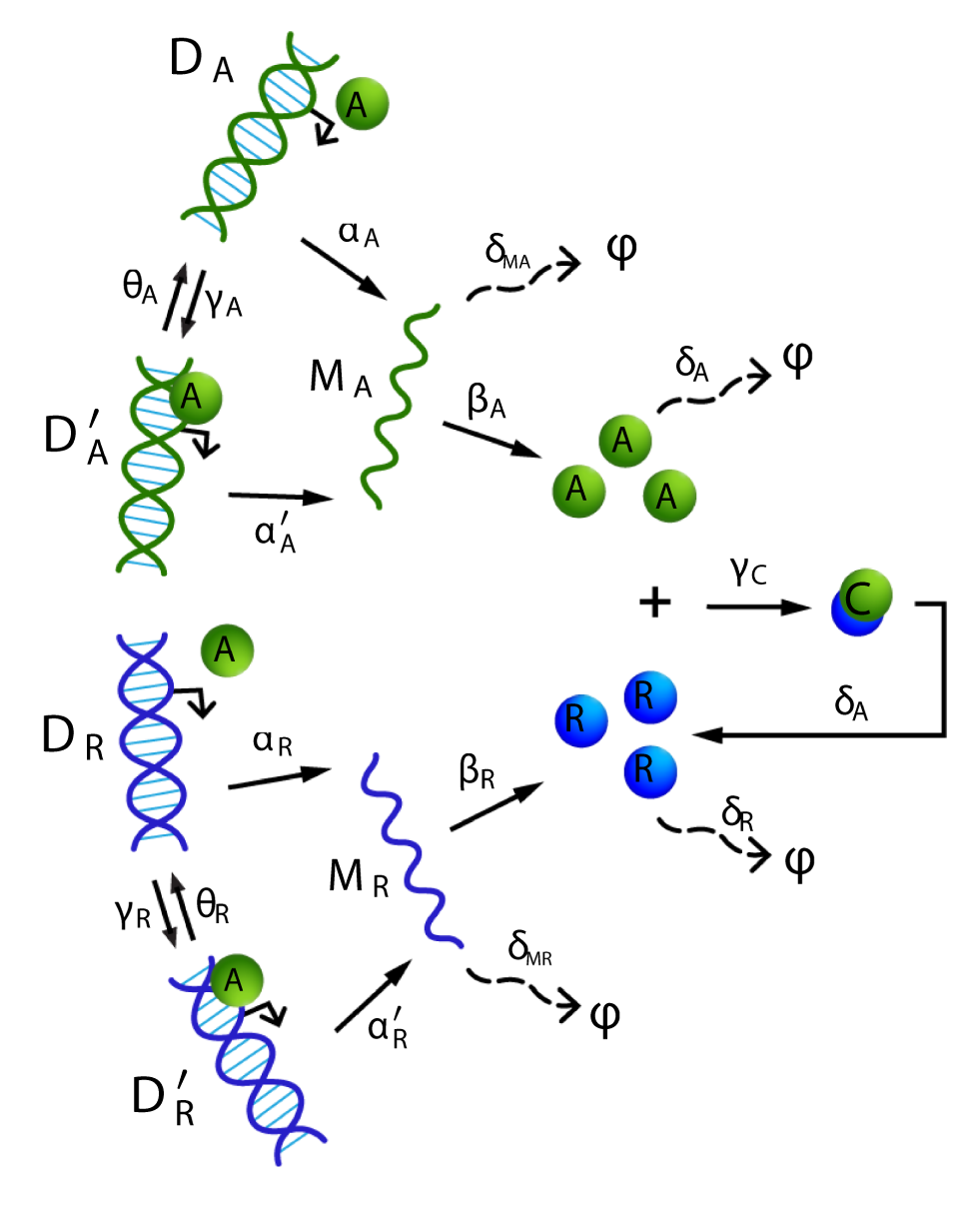}
  \caption{\textbf{Circadian clock model of \cite{Vilar:02}.}}\label{fig:clock}
\end{figure}

Let us consider the circadian oscillator of \cite{Vilar:02}, depicted in Figure \ref{fig:clock}, which is a network involving 9 species and 18 reactions.

\begin{figure}[!ht]
    \centering
    \includegraphics[width=0.75\textwidth]{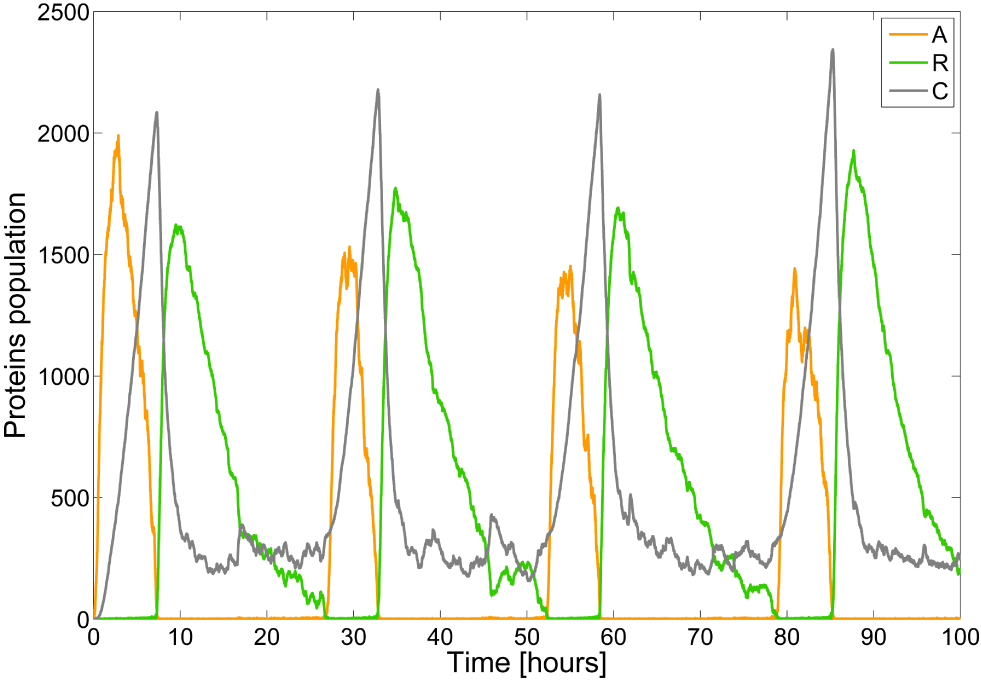}
    \caption{\textbf{Sample-path of the species of the circadian clock model.}}\label{fig:clock_SP}
\end{figure}

\begin{figure*}[!ht]
\centering
    \includegraphics[width=0.95\textwidth]{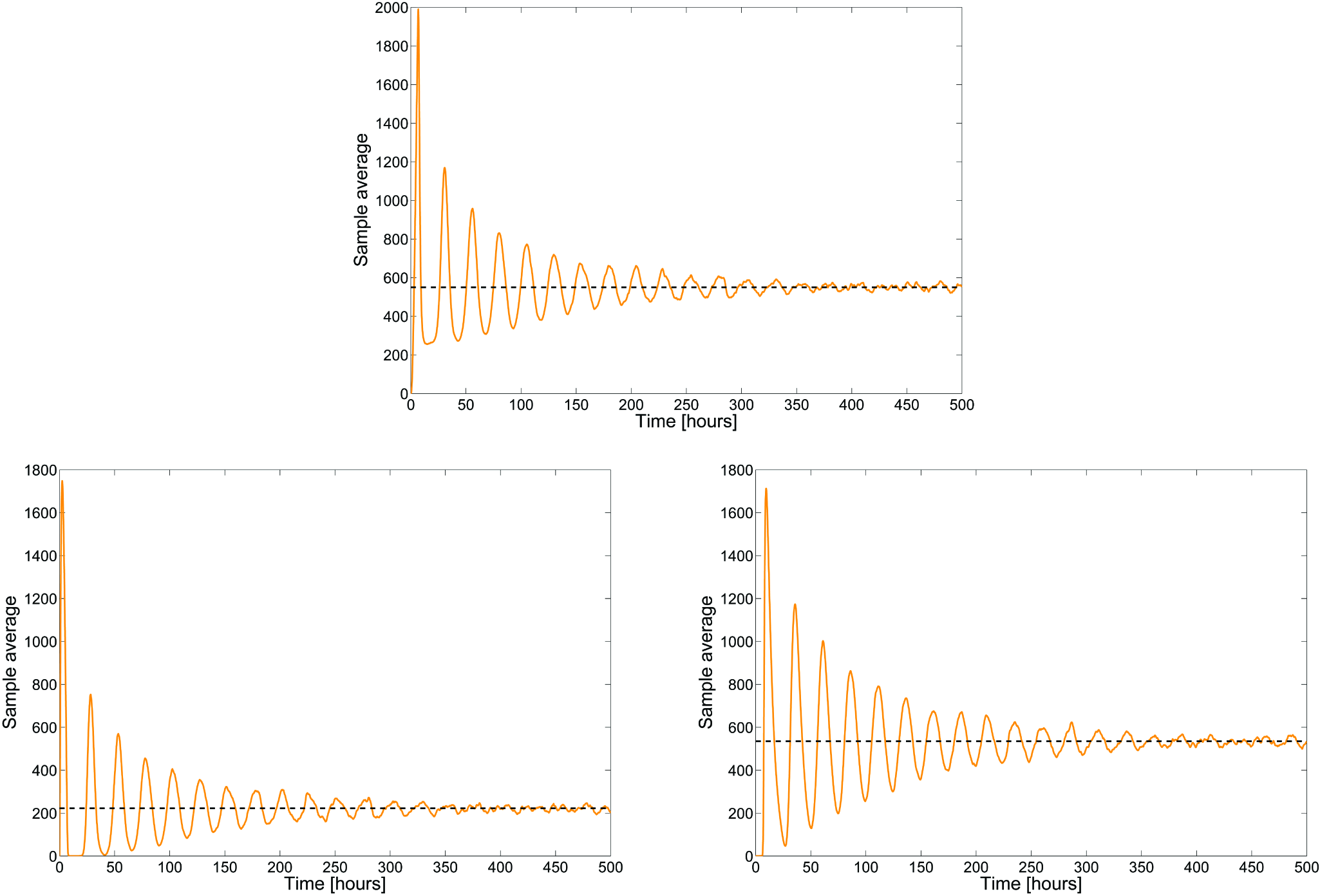}
  \caption{\textbf{Time evolution of the sample averages of the species \bf{A} (top), \bf{R} (left) and \bf{C} (right) of the circadian clock model (2000 cells averaging)}.  The dashed-lines correspond to the (asymptotic) time-average.}\label{fig:SATA}
\end{figure*}

%
%
Applying the developed theory on this model, we obtain the following result:
  \begin{result}
  For any positive values of the rate parameters, the circadian clock model  of \cite{Vilar:02} is ergodic and all the moments are bounded and globally converging.
\end{result}
Using, for instance, the values of \cite{Vilar:02} and solving for the optimization problem \eqref{eq:optlf3} using linprog  and Yalmip \cite{YALMIP}, we find that $c_1=402.5768$ and $c_2=0.1992$. Typical trajectories for the proteins \textbf{A}, \textbf{R} and \textbf{C} are depicted in Figure~\ref{fig:clock_SP} where we can observe the expected oscillatory behavior. When averaging the populations of the proteins \textbf{A}, \textbf{R} and \textbf{C} over a population of 2000 cells, we obtain the sample-average trajectories depicted in Figure~\ref{fig:SATA}. Convergence to stationary values is easily seen. Moreover, from the ergodicity property, we can even state that these fixed points for the sample-averages are globally attracting and that they coincide with the asymptotic time-average (dashed lines). The steady-state average values for the proteins \textbf{A}, \textbf{R} and \textbf{C} are given by 222.1797, 534.8853 and 549.7195, respectively.

\subsection*{p53 model}

Let us consider one of the oscillatory p53 models of \cite{Geva:06}, which is described by the reactions
\begin{equation}\label{eq:p53}
  \begin{array}{rclrclrcl}
    \phib &\stackrel{k_1}{\longrightharpoonup}&\X{1}, &   \X{1}&\stackrel{k_2}{\longrightharpoonup}&\phib ,&   \X{1}&\stackrel{{\small f(\X{1},\X{3})}}{\longrightharpoonup}&\phib \\
   \X{3}&\stackrel{k_{6}}{\longrightharpoonup}&\phib  ,&    \X{2}&\stackrel{k_5}{\longrightharpoonup}&\X{3},&
    \X{1}&\stackrel{k_4}{\longrightharpoonup}&\X{1}+\X{2}.
    \end{array}
\end{equation}
where $\X{1}$ is the number of p53 molecules, $\X{2}$ the number of precursor of Mdm2 molecules and $\X{3}$ the number of molecules of Mdm2. The function $\textstyle f(x,y)=\frac{k_3y}{x+k_7}$ implements a nonlinear feedback on the degradation rate of p53. We have the following result:
\begin{result}
  For any positive values of the rate parameters, the oscillatory p53 model \eqref{eq:p53} is ergodic and all the moments are bounded and globally converging.
\end{result}

\subsection*{Lotka-Volterra model}

We consider here the stochastic reaction network
\begin{equation}\label{scrn:volterra}
  \begin{array}{rclcrcl}
    \phib &\stackrel{\alpha_i}{\longrightharpoonup}&\X{i},&&  \X{i}&\stackrel{\beta_i}{\longrightharpoonup}&\X{i}+\X{i}\\
    \X{i}+\X{j}&\stackrel{\gamma_{ij}}{\longrightharpoonup}&\X{j},&&    \X{i}&\stackrel{\delta_i}{\longrightharpoonup}&\phib
  \end{array}
\end{equation}
which is an open analogue of the deterministic Lotka-Volterra system of \cite{Gopalsamy:84}. 
The first set of reactions represent immigration, the second one reproduction, the third one competition due to overpopulation and the last one deaths/migrations. We obtain then the following result, which is a stochastic analogue of the results in \cite{Champagnat:10} obtained in the deterministic setting:
\begin{theorem}
   Let us define $\Gamma(v)=[v_i\gamma_{ij}]$
 and assume that one of the following conditions hold:
  \begin{enumerate}
    \item there exists $v>0$ such that the matrix $\Gamma(v)+\Gamma(v)^\T$ is positive definite;
    \item there exists $v>0$ such that the $\Gamma(v)+\Gamma(v)^\T$ is copositive, i.e. $x^T(\Gamma(v)+\Gamma(v)^\T)x\ge0$ for all $x\ge0$, and $\beta_i-\delta_i<0$ for all $i=1,\ldots,n$.
  \end{enumerate}
  Then, the stochastic reaction network \eqref{scrn:volterra} is ergodic and all the moments up to order $\left\lfloor1+\dfrac{2c_2}{c_5}\right\rfloor-2$ are bounded and globally converging.
\end{theorem}

\red{\subsection*{Schl\"{o}gl model}}

\red{In order to illustrate that the method can be applied to systems with more general mass-action kinetics, we consider the stochastic version of the well-known Schl\"{o}gl model \cite{Schlogl:72}:
\begin{equation}\label{eq:schlogl}
  \begin{array}{rclcl}
    2\Xo&\rarrow{k_1X_A}&3\Xo&\rarrow{k_2}&2\Xo\\
    \phib&\rarrow{k_3}&\Xo&\rarrow{k_4 X_B}&\phib
  \end{array}
\end{equation}
where $\Xo$ is the main molecule in the network. The above model is derived in the supplementary material \textbf{S1} where we have assumed that the other molecular populations do not vary over time. Note that in the present form the model has an infinite state-space and involves a single trimolecular reaction. We then have the following result.}

\red{\begin{theorem}
  For any positive values of the rate parameters $k_1,k_2,k_3,k_4$ and any positive values for $X_A$ and $X_B$, the Markov process describing the Schl\"{o}gl model \eqref{eq:schlogl} is exponentially ergodic.
\end{theorem}}

\red{Note, however, that we cannot say anything on the stability of the moments (besides the fact that the first order-moment converges) since the condition {\bf DD2} does not hold here due to the presence of a cubic term. Note that extending the condition \textbf{DD2} to handle more general cases, such as this one, might be possible.}

\section*{Discussion}

The central theme of this paper is to verify the ergodicity and moment boundedness of reaction networks in the stochastic setting. Note that even though we mainly consider mass-action kinetics in this paper, the framework also applies to more general kinetics described, for instance, by Hill functions (see the examples on the repressilator and the stochastic switch) \red{and more general mass-action kinetics}. These results have several interesting and important biological implications.

For example, the ergodicity of a network shows that population-level information could be obtained by observing a single trajectory for a long time. Such an insight can be used to leverage different experimental techniques for a
given application. For example, consider a clonal cell population with each cell having a gene-expression
network that is ergodic. Then the stationary distribution (at the population level) of the species involved in this network can be ascertained by {\em observing a single cell} over time. In other words, to obtain stationary distributions
one can either collect samples over time from a single cell (e.g. using time-lapse microscopy) or one can take a snapshot of the entire cell population at some fixed time (e.g. using flow-cytometry). Due to ergodicity, both these approaches will yield the same information. Hence, far from being a technical condition, ergodicity can have far reaching experimental implications.

\begin{figure}[!ht]
  \centering
  \includegraphics[width=0.8\textwidth]{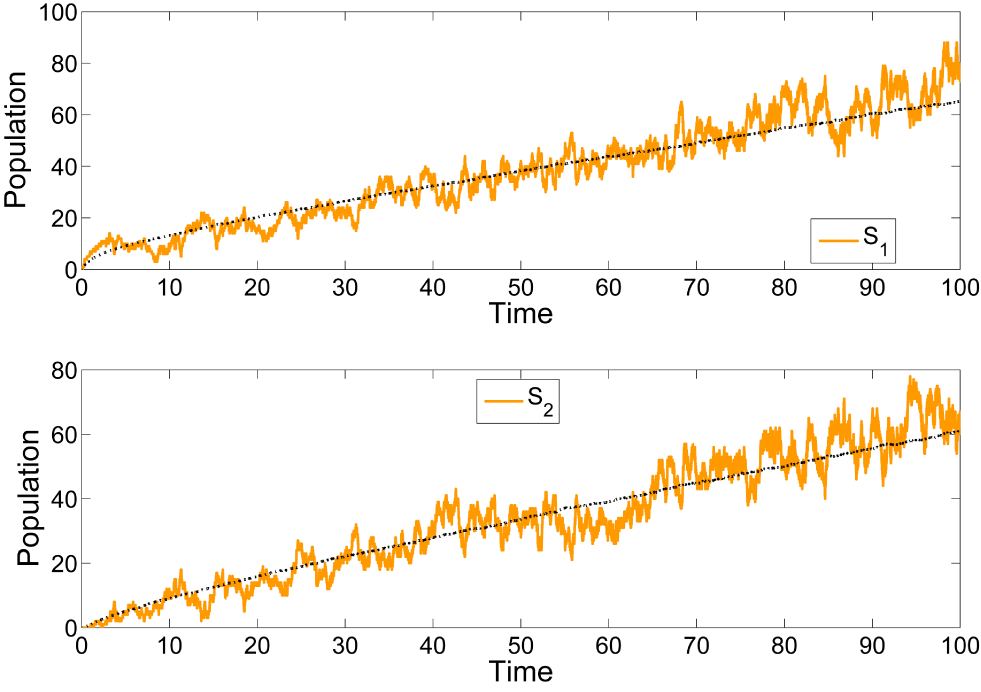}
  \caption{\textbf{State trajectories of the carcinogenesis model \eqref{scrn:cancer} with the parameters $k_1=5$, $k_{12}=1$, $k_{21}=1$, $\gamma_2=4$ and $\alpha=1$.} The dashed lines correspond to the average trajectories computed over 1000 cells.}\label{fig:cancer}
\end{figure}

As a property of a network, ergodicity also sheds important light on the long range behaviors that can be exhibited by that network. One may expect that most endogenous  biochemical networks to be ergodic in order to achieve robustness with respect to variability in initial conditions and  kinetic parameters, thus ensuring proper biological functions in spite of environmental disturbances. As also mentioned in the introduction, ergodicity is a non-trivial property which needs to be carefully established and cannot be generically assumed. To illustrate this, let us consider a simplified version of the model of carcinogenesis considered in \cite{Bois:90} which is given by

\begin{equation}\label{scrn:cancer}
  \begin{array}{rclcrcl}
    \phib &\stackrel{k_1}{\longrightharpoonup}&\X{1},&&  \X{1}&\stackrel{k_{12}}{\longrightharpoonup}&\X{2}\\
    \X{2}&\stackrel{k_{21}}{\longrightharpoonup}&\X{1},&&    \X{2}&\stackrel{f(x)}{\longrightharpoonup}&\phib
  \end{array}
\end{equation}
where $f(x)=\dfrac{\gamma_2}{\alpha+x_2}$, $\alpha>0$. When $k_1>\gamma_2$, the trajectories of the species grow unbounded, as shown in Figure~\ref{fig:cancer}, emphasizing then non-ergodicity of the model for this choice of parameters.

The ideas we use for analysis can also be applied for rationally designing circuits in synthetic
biology, where it is important that the network be (structurally) ergodic in order to ensure that the
dynamics has the desired behavior irrespective of the initial conditions. Such a design is crucial because
the initial conditions are usually unknown or difficult to control at certain times, e.g. after cell division or
after the transfection of plasmids in the cell.

Our results on boundedness and convergence of statistical moments enable verification of the suitability
of a stochastic model and to characterize the properties of its steady-state distributions, even if such a distribution is not
explicitly computable.  One application of this is to provide justifications and insights for using moment
closure techniques which have been extensively used to study stochastic chemical reaction networks.
Some of these techniques \cite{Lee:09,Ale:13} are based on manipulations of the moment generating function of the
underlying stochastic process. The existence of this moment generating function is implicitly assumed in
such techniques but it may not always hold, thereby jeopardizing the validity of the technique. In this
article, we show that under certain conditions, the distribution of the stochastic process is uniformly light-tailed,
which proves that the moment generating function exists for all time. Certain moment closure
techniques (see \cite{Gomez:07,Singh:11}) prescribe ways to approximate higher order moments as a function of lower order
moments. Such an approximation is, however, only reasonable if the higher order moments are bounded
over time. This can be easily assessed with our approach and one can even quantify the error by explicitly
computing the moment bounds as described in this article.

Finally, the techniques developed here will prove invaluable for designing synthetic biological control systems and
circuits  whose objective is to steer the moments of the network of interest to a specific steady-state value. Until now, no theory
has provided guidance for such a design. The specifics are outside the scope of this article and will be pursued elsewhere.

\section*{Acknowledgments}

The authors are grateful to Stephanie Aoki and Christine Khammash who spent some of their precious time in producing several illustrative pictures.

\section*{Author contributions}

C.B. and A.G. contributed equally to this work. C.B., A.G. and M.K. devised the research; C.B. and A.G. carried out the research; C.B., A.G. and M.K. wrote the paper; A.G. developed the mathematical framework; C.B. developed the results for \red{unimolecular and bimolecular reaction networks}, and applied them to the examples.

\section*{Funding}

This work has been supported by ETH and  the Human Frontier Science Program Grant RGP0061/2011.

\bibliographystyle{plos2009}
















%

\end{document}